\overfullrule=0pt
\input harvmac
\def\wh{\hat}

\def\hth{\hat\theta}
\def\hd{\hat d}

\def\hl{\hat\lambda}
\def\hht{\hat\theta}

\def\a{{\alpha}}
\def\ah{{\hat\alpha}}

\def\l{{\lambda}}

\def\b{{\beta}}
\def\bh{{\hat\beta}}
\def\g{{\gamma}}
\def\k{{\kappa}}
\def\gh{{\hat\gamma}}
\def\d{{\delta}}
\def\dh{{\hat\delta}}
\def\kh{{\hat\kappa}}
\def\e{{\epsilon}}
\def\s{{\sigma}}

\def\L{{\Lambda}}

\def\O{{\Omega}}
\def\hO{{\hat\Omega}}

\def\half{{1\over 2}}
\def\p{{\partial}}

\def\t{{\theta}}

\Title{\vbox{\hbox{IFT-P.063/2002 }}}
{\vbox{
\centerline{\bf ICTP Lectures on Covariant Quantization of the Superstring}}}
\bigskip
\centerline{Nathan Berkovits\foot{e-mail: nberkovi@ift.unesp.br}}
\bigskip
\centerline{\it Instituto de F\'\i sica Te\'orica, Universidade Estadual
Paulista}
\centerline{\it Rua Pamplona 145, 01405-900, S\~ao Paulo, SP, Brasil}

\vskip .3in

These ICTP Trieste lecture notes review the pure spinor approach to quantizing
the superstring with manifest
D=10 super-Poincar\'e invariance. The first section discusses
covariant quantization of the superparticle and gives a new
proof of equivalence with the Brink-Schwarz superparticle. The
second section discusses the superstring in a flat background and
shows how to construct vertex operators and compute tree amplitudes
in a manifestly super-Poincar\'e covariant manner. And the third
section discusses quantization of 
the superstring in curved backgrounds which
can include Ramond-Ramond flux. 

\Date {September 2002}

\newsec{Introduction}

The two standard formalisms for describing the superstring are the
Ramond-Neveu-Schwarz (RNS) and Green-Schwarz (GS) formalisms.
Although the RNS formalism has a beautiful N=1 worldsheet supersymmetry,
its lack of manifest target-space supersymmetry is responsible for several
awkward features of the formalism. For example, amplitudes involving
more than four external fermions are difficult to compute in a
Lorentz-covariant manner because of picture-changing and bosonization
complications \ref\sixf{V.A. Kostelecky, O. Lechtenfeld, S. Samuel,
D. Verstegen, S. Watamura and D. Sahdev, {\it The Six Fermion
Amplitude in the Superstring}, Phys. Lett. B183 (1987) 299.}. 
Furthermore, it is not known how to use the RNS
formalism to describe the superstring in Ramond-Ramond backgrounds.

On the other hand, target-space supersymmetry is manifest in the GS
formalism, but the worldsheet symmetries are not manifest.
A lack of understanding of these worldsheet symmetries has so far
prevented quantization except in light-cone gauge. Although
light-cone gauge is useful for determining the physical spectrum,
it is clumsy for computing scattering amplitudes because
of the lack of manifest Lorentz covariance and the need to
introduce interaction-point operators and contact terms. For
these reasons, only four-point tree and one-loop amplitudes have
been explicitly computed using the GS formalism \ref\dualgs{M.B. Green
and J.H. Schwarz, {\it Supersymmetrical Dual String Theory}, Nucl. Phys.
B181 (1981) 502.}.
Furthermore, the necessity of choosing light-cone gauge means that
quantization is only possible in those backgrounds which allow
a light-cone gauge choice.

As will be discussed in these lecture notes, a new formalism for
the superstring was proposed recently \ref\cov{N. Berkovits,
{\it Super-Poincar\'e Covariant Quantization of the Superstring},
JHEP 0004 (2000) 018, hep-th/0001035.} which combines the advantages
of the RNS and GS formalisms without including their disadvantages.
In this new approach, the worldsheet action is quadratic in a flat
background so quantization is as easy as in the RNS formalism.
And since D=10 super-Poincar\'e covariance is manifest in this formalism,
there is no problem with computing spacetime-supersymmetric N-point
tree amplitudes or with quantizing the superstring in Ramond-Ramond
backgrounds. 

There are three new ingredients in this formalism as compared with
the standard GS formalism. The first new ingredient is fermionic
canonical momenta $d_\a$ for the $\t^\a$ variables. These canonical
momenta were first introduced by Siegel \ref\siegelcs{W. Siegel,
{\it Classical Superstring Mechanics}, Nucl. Phys. B263 (1986) 93.}
and allow the GS action to be written in quadratic form after including
appropriate constraints. The second new ingredient is the bosonic
``pure spinor'' $\l^\a$ which plays the role of a ghost variable.
And the third new ingredient is the nilpotent
BRST operator $Q=\int\l^\a d_\a$ whose cohomology is used to define
physical states. But before entering into more details about this new
formalism, it will be useful to say a few words about where it
came from.

In 1989, in an attempt to better understand the worldsheet symmetries
of the GS superstring, Sorokin, Tkach, Volkov and Zheltukhin \ref\kharkov
{D.P. Sorokin, V.I. Tkach, D.V. Volkov and A.A. Zheltukhin,
{\it From the Superparticle Siegel Symmetry to the Spinning Particle
Proper Time Supersymmetry}, Phys. Lett. B216 (1989) 302.}
replaced the worldline kappa symmetry of the Brink-Schwarz
superparticle with worldline supersymmetry.
The bosonic worldline
superpartner for $\t^\a$ was called $\l^\a$, and worldline
supersymmetry of the action implied that $\l^\a$ satisfied the
twistor-like relation 
\eqn\twistor{\l\g^m\l = \dot x^m +{1\over 2} \t\g^m \dot\t.}
This twistor-like approach was then generalized by several authors 
to the classical heterotic superstring with from one to eight
worldsheet supersymmetries \ref\hetone{N. Berkovits,
{\it A Covariant Action for the Heterotic Superstring with Manifest
Spacetime Supersymmetry and Worldsheet Superconformal Invariance},
Phys. Lett. B232 (1989) 184.}\ref\hettwo{M. Tonin, {\it 
Worldsheet Supersymmetric
Formulations of Green-Schwarz Superstrings}, Phys. Lett. B266 (1991)
312.} \ref\heteight{
F. Delduc, A. Galperin, P.S. Howe and E. Sokatchev,
{\it A Twistor Formulation of the Heterotic D=10 Superstring with Manifest
(8,0) Worldsheet Supersymmetry}, Phys. Rev. D47 (1993) 578, hep-th/9207050.}
and it was argued in \ref\hetme{N. Berkovits, {\it The Heterotic
Green-Schwarz Superstring on an N=(2,0) Worldsheet}, Nucl. Phys. B379
(1992) 96, hep-th/9201004.}
that quantization of the version with two worldsheet supersymmetries
leads to a critical N=2 superconformal field theory.
For two worldsheet supersymmetries, $\t^\a$ has two superpartners,
$\l^\a$ and $\bar\l^\a$, which satisfy the relations
\eqn\twistortwo{\l\g^m\l=\bar\l\g^m\bar\l=0,\quad
\l\g^m\bar\l = \p x^m +{1\over 2} \t\g^m\p\t.}
In ten dimensions, a complex Weyl spinor $\l^\a$ satisfying
$\l\g^m\l=0$ is called a pure spinor and, as was shown by Howe 
\ref\howeone{P.S. Howe, {\it Pure Spinor Lines in Superspace and
Ten-Dimensional Supersymmetric Theories}, Phys. Lett. B258 (1991) 141.}
\ref\howetwo{
P.S. Howe, {\it Pure Spinors, Function Superspaces and Supergravity
Theories in Ten Dimensions and Eleven Dimensions}, Phys. Lett. B273 (1991)
90.} 
in 1991,
is useful for describing the on-shell constraints of 
super-Yang-Mills and supergravity.\foot{Pure spinors were originally
studied by Cartan \ref\cartan{E. Cartan,{\it Lecons sur la Theorie des
Spineurs}, Hermann, Paris, 1937.}. 
They have also been used for defining grand
unified models \ref\paulo{P. Budinich, {\it From the Geometry of Pure
Spinors with their Division Algebras to Fermion's Physics},
hep-th/0107158\semi P. Budinich and A. Trautman, {\it
Fock Space Description of Simple Spinors}, J. Math. Phys. 30 (1989) 2125.}
and for constructing super-Yang-Mills auxiliary fields \ref\nilsson{
B.E.W. Nilsson, {\it Pure Spinors as Auxiliary Fields in the 
Ten-Dimensional Supersymmetric Yang-Mills Theory}, Class. Quant. Grav. 3
(1986) L41.}.}

Unfortunately, direct quantization of the 
N=2 worldsheet superconformal
field theory
requires solving the constraints of \twistortwo\ and breaking
the manifest SO(9,1) Lorentz invariance down to U(4) \hetme\ref\twistme
{N. Berkovits, {\it Calculation of Green-Schwarz Superstring Amplitudes
using the N=2 Twistor-String Formalism}, Nucl. Phys. B395 (1993) 77,
hep-th/9208035.}. 
In later papers, this U(4) formalism was related to other critical
N=2 superconformal field theories called ``hybrid'' formalisms
with manifest SO(3,1)$\times$U(3) \ref\fourme
{N. Berkovits, {\it Covariant Quantization of the Green-Schwarz
Superstring in a Calabi-Yau Background}, Nucl. Phys. B431 (1994) 258,
hep-th/9404162.},
SO(5,1)$\times$U(2) \ref\topo{N. Berkovits and C. Vafa,
{\it N=4 Topological Strings},
Nucl. Phys. B433 (1995) 123, hep-th/9407190.},
SO(1,1)$\times$U(4) \ref\valtwo{N. Berkovits, 
S. Gukov and B.C. Vallilo,
{\it Superstrings in 2D Backgrounds with R-R Flux and New Extremal
Black Holes}, Nucl. Phys. B614 (2001) 195, hep-th/0107140.},
or (after Wick-rotation) U(5) \ref\ufive{N. Berkovits, {\it Quantization 
of the Superstring with Manifest
U(5) Super-Poincar\'e Invariance}, Phys. Lett. B457 (1999) 94, 
hep-th/9902099.}
subgroups of the Lorentz group. Together with Cumrun Vafa
\ref\unif{N. Berkovits, {\it The 
Ten-Dimensional Green-Schwarz Superstring is a
Twisted Neveu-Schwarz-Ramond String}, Nucl. Phys. B420 (1994) 332,
hep-th/9308129\semi
N. Berkovits and C. Vafa, {\it On the Uniqueness of String Theory},
Mod. Phys. Lett. A9 (1994) 653, hep-th/9310170.}\topo,
it was
shown that all of these formalisms are related by a field
redefinition to an N=1 $\to$ N=2 embedding of the standard
RNS formalism where, after twisting the worldsheet N=2, the
RNS BRST current and $b$ ghost are mapped to the 
fermionic N=2 superconformal generators. 

Finally, in \cov, it was proposed that
these hybrid formalisms are equivalent to a manifestly
SO(9,1) super-Poincar\'e covariant formalism using a BRST
operator $Q=\int \l^\a d_\a$ constructed
from the worldsheet variables $[x^m, \t^\a, d_\a,\l^\a, w_\a]$
where $d_\a$ is the conjugate momentum to $\t^\a$, $w_\a$ is
the conjugate momentum to $\l^\a$, and $\l^\a$ is a pure
spinor satisfying $\l\g^m\l=0$. As will be shown later,
$\l^\a$ and $w_\a$ each contain 11 independent components so the
covariant formalism contains 32 bosons and 32 fermions.
Since the hybrid formalisms all
contain 12 bosons and 12 fermions (which are related by a field
redefinition to the RNS variables $[x^m,\psi^m, b,c,\beta,\gamma]$),
the proposal is based on the conjecture that, in addition to
obeying the usual physical state conditions, states in the cohomology
of $Q=\int \l^\a d_\a$ are independent of the extra 20 bosons and 20 fermions.

This conjecture was suggested by the U(5) version \ufive\
of the hybrid formalism
whose variables are $[x^m,\t^a,\t^+,d_a,d_+,\l^+,w_+]$ where $a=1$ to 5.
If $\l^\a=\l^+$ is interpreted as choosing a U(5) direction in
SO(10), the extra 20 bosons can be understood as parameterizing the
SO(10)/U(5) coset space. In this sense, the projective part of
the pure spinor variable plays the role of an SO(10)/U(5)
harmonic variable, similar to the attempts of \ref\harmattempts
{E. Sokatchev, {\it Harmonic Superparticle}, Class. Quant. Grav. 4 (1987) 
237\semi E.R. Nissimov and S. J. Pacheva, {\it Manifestly Super-Poincar\'e 
Covariant Quantization of the Green-Schwarz Superstring},
Phys. Lett. B202 (1988) 325\semi R. Kallosh and M. Rakhmanov,
{\it Covariant Quantization of the Green-Schwarz
Superstring}, Phys. Lett. B209 (1988) 233.} to covariantly quantize
the superstring.

After the proposal was made in \cov, there have been various
consistency checks of its validity. These include a proof that
the cohomology of $Q=\int\l^\a d_\a$ reproduces the  
superstring spectrum \ref\cohomo{N. Berkovits,
{\it Cohomology in the Pure Spinor Formalism for the
Superstring}, JHEP 0009 (2000) 046, hep-th/0006003\semi
N. Berkovits and O. Chand\'{\i}a,
{\it Lorentz Invariance of the Pure Spinor
BRST Cohomology for the Superstring},
Phys. Lett. B514 (2001) 394, hep-th/0105149.}
and the construction of an explicit
map from states in the cohomology of $Q$ to physical states in the
RNS formalism \ref\rnsmap{N. Berkovits,
{\it Relating the RNS and Pure Spinor Formalisms for the
Superstring}, JHEP 0108 (2001) 026, hep-th/0104247.}.
Also, the pure spinor description has been
generalized to curved backgrounds and it has been shown that
BRST invariance implies the correct low-energy equations of motion
for the background fields \ref\born{N. Berkovits and
V. Pershin, {\it Supersymmetric Born-Infeld from the Pure Spinor
Formalism of the Superstring}, hep-th/0205154.}\ref\withhowe{N. Berkovits
and P. Howe, {\it Ten-Dimensional Supergravity Constraints from the
Pure Spinor Formalism for the Superstring}, Nucl. Phys. B635 (2002) 75,
hep-th/0112160.}.
Furthermore, it has recently been shown (at 
least at the
classical level) that the pure spinor description can be obtained by
directly gauge-fixing the original N=2 worldsheet supersymmetric
description \hettwo\hetme\ 
of \twistortwo\ without passing through the hybrid or RNS descriptions
\ref\toninnew{M. Matone, L. Mazzucato, I. Oda, D. Sorokin and
M. Tonin, {\it The Superembedding Origin of the Berkovits Pure Spinor
Covariant Quantization of Superstrings}, Nucl. Phys. B639 (2002) 182,
hep-th/0206104.}.

Although on-shell states in the pure spinor description can be
related to on-shell states in the RNS description \rnsmap,
there is no such relation for off-shell states. Note that
the super-Poincar\'e algebra
closes for both on-shell and off-shell states in the pure
spinor description. But in the RNS 
descriptions, the super-Poincar\'e algebra closes up to
picture-changing \ref\fms
{D. Friedan, E. Martinec and S. Shenker,
{\it Conformal Invariance, Supersymmetry and String
Theory}, Nucl. Phys. B271 (1986) 93.}, 
which is only defined for on-shell states. Since
there is no off-shell map between the descriptions, it is tricky
to guess
the correct rules for computing scattering
amplitudes. Nevertheless, a manifestly super-Poincar\'e covariant
prescription was given for tree amplitudes using the pure
spinor description and was shown in \ref\ampval{N. Berkovits and
B.C. Vallilo, {\it Consistency of Super-Poincar\'e Covariant Superstring
Tree Amplitudes}, JHEP 0007 (2000) 015, hep-th/0004171.}\rnsmap\
to coincide with the RNS prescription. However, it is still unknown
how to compute manifestly super-Poincar\'e covariant loop
amplitudes using the pure spinor description. 
It is possible that recent generalizations of the
pure spinor approach which explicitly
introduce $[b,c]$ reparameterization ghosts
may be useful for defining a loop amplitude prescription \ref\vann
{P.A. Grassi, G. Policastro, M. Porrati and P. Van Nieuwenhuizen,
{\it Covariant Quantization of Superstrings without Pure Spinor
Constraints}, hep-th/0112162\semi
P.A. Grassi, G. Policastro and P. Van Nieuwenhuizen,
{\it The Massless Spectrum of Covariant Superstrings}, hep-th/0202123\semi
P.A. Grassi, G. Policastro and P. Van Nieuwenhuizen,
{\it On the BRST Cohomology of Superstrings with/without Pure Spinors},
hep-th/0206216\semi
P.A. Grassi, G. Policastro and P. Van Nieuwenhuizen,
{\it The Covariant Quantum Superstring and Superparticle from their
Classical Actions}, hep-th/0209026.}\ref\vaman{R. Roiban, W. Siegel
and D. Vaman, private communication.}.

In section 2 of these notes, covariant quantization of the superparticle
using pure spinors will be reviewed and a previously
unpublished proof will be given for
equivalence with the Brink-Schwarz superparticle.
In section 3, the pure spinor approach will be generalized to the
superstring and it will be shown how to construct massless and massive 
vertex operators and compute tree amplitudes in a manifestly 
super-Poincar\'e covariant manner.
In section 4, the open and closed 
superstring will be described in a curved background
and it will be shown how to obtain the low-energy supersymmetric
Born-Infeld and supergravity
equations of motion
for the background fields from the condition of BRST invariance.
It will also be shown how this approach can be used to
quantize the superstring in an $AdS_5\times S^5$ background (or
its plane wave limit) with Ramond-Ramond flux.\foot{Some material
in this review, such as massive vertex operators and supersymmetric
Born-Infeld, were not included in the ICTP lectures. Also, the
lecture on quantization of the d=11 superparticle and supermembrane
was not included in this review since it involves work in progress.}

\newsec{Covariant Quantization of the Superparticle}

Before discussing the pure spinor description, it will be useful
to review the standard description of the superparticle and the
superspace equations for ten-dimensional super-Yang-Mills. It will
then be shown that just as D=3 Chern-Simons theory can be obtained
from BRST quantization of a particle action, D=10 super-Yang-Mills
theory can be obtained from BRST quantization of a superparticle
action involving pure spinors.

\subsec{Review of standard superparticle description}

The standard Brink-Schwarz
action for the ten-dimensional superparticle is 
\ref\super{L. Brink and J.H. Schwarz, {\it Quantum Superspace},
Phys. Lett. B100 (1981) 310\semi A. Ferber, {\it Supertwistors
and Conformal Supersymmetry}, Nucl. Phys. B132 (1978) 55.}
\eqn\action{
S=\int d\tau (\Pi^m P_m + e P^m P_m)}
where 
\eqn\defpi{\Pi^m = \dot x^m - {1\over 2} \dot\t^\a\g^m_{\a\b}\t^\b,}
$P_m$ is the canonical momentum for $x^m$, and $e$
is the Lagrange multiplier which enforces the mass-shell condition.
The gamma matrices $\g^m_{\a\b}$ and $\g_m^{\a\b}$ are
$16\times 16$ symmetric matrices which satisfy
$\g^{(m}_{\a\b} \g^{n)~ \b\g}= 2 \eta^{mn} \d_\a^\g$.
In the Weyl representation, 
$\g^m_{\a\b}$ and $\g_m^{\a\b}$ are
the off-diagonal blocks of the $32\times 32$ $\Gamma^m$ matrices.
Throughout these notes, the conventions for factors of $i$ and 2 will
be chosen such that the supersymmetry algebra is $\{q_\a,q_\b\}=
\g^m_{\a\b} \p_m =i P_m \g^m_{\a\b}$.

The action of \action\ is spacetime-supersymmetric under
$$\d \t^\a = \e^\a, \quad x^m = {1\over 2}\t\g^m\e, \quad \d P_m = \d e =0,$$
and is also invariant under the local $\kappa$ transformations
\ref\sieg{W. Siegel, {\it Hidden Local Supersymmetry in the
Supersymmetric Particle Action}, Phys. Lett. B128 (1983) 397.}
\eqn\ka{ \d \t^\a = P^m (\g_m \k)^\a,\quad
\d x^m = -{1\over 2}\t\g^m\d\t ,\quad \d P_m =0,\quad \d e= \dot\t^\b\k_\b.}
The canonical momentum to $\t^\a$, which will be called $p_\a$, satisfies
$$ p_\a = \d L/ \d\dot\t^\a =-{1\over 2} P^m (\g_m\t)_\a,$$
so
canonical
quantization requires that physical states are annihilated by the
fermionic Dirac constraints defined by
\eqn\dirac{d_\a = p_\a +{1\over 2} P_m (\g^m\t)_\a.}
Since $\{p_\a,\t^\b\}=-i\d_\a^\b$,
these constraints satisfy the Poisson brackets 
\eqn\antic{\{d_\a, d_\b\} =-i  P_m \g^m_{\a\b},}
and since $P^m P_m =0$ is also a
constraint, eight of the sixteen Dirac constraints
are first-class and eight are second-class.
One can easily check that the eight first-class Dirac constraints
generate the $\kappa$ transformations of \ka, however, there
is no simple way to covariantly separate out the second-class
constraints. 

Nevertheless, one can easily quantize the superparticle in a non-Lorentz
covariant manner and obtain the physical spectrum.
Assuming non-zero $P^+$,
the local fermionic $\k$-transformations can be used to
gauge-fix $(\g^+\t)_\a =0$ where
$\g^\pm = {1\over{\sqrt{2}}}
(\g^0 \pm \g^9)$. In this ``semi-light-cone'' gauge, the action of \action\
simplifies to the quadratic action 
\eqn\lcactionone{S=\int d\tau (\dot x^m P_m +{1\over 2}
 P^+ (\dot\t \g^- \t) + e P^m P_m)}
\eqn\lcaction{
=
\int d\tau (\dot x^m P_m + {1\over 2}\dot S_a S_a + e P^m P_m) ,}
where $S_a = \sqrt{P^+} (\g^-\t)_a$ and
$a=1$ to 8 is an $SO(8)$ chiral spinor index.

Canonical quantization of \lcaction\ implies that
$\{S_a, S_b \} = i\d_{ab}$. So $S_a$ 
acts like a `spinor' version of
$SO(8)$ Pauli matrices
$\s^j_{a\dot b}$ which are normalized to satisfy
$$\s^j_{a \dot c} 
\s^j_{b \dot d} + 
\s^j_{b \dot c} 
\s^j_{a \dot d} = i \d_{ab} \d_{\dot c \dot d}$$ 
where $j$ and $\dot b$ are $SO(8)$ vector and antichiral spinor indices.
One can therefore define the quantum-mechanical wavefunction $\Psi(x)$
to carry either an $SO(8)$ vector index, $\Psi_j(x)$,
or an $SO(8)$ antichiral spinor index, $\Psi_{\dot a}(x)$, and the
anticommutation relations of $S_a$ are reproduced by defining
\eqn\wfc{S^a \Psi_j(x) = \sigma_j^{a\dot b} \Psi_{\dot b}(x),\quad
S_a \Psi_{\dot b}(x) = \sigma^j_{a\dot b} \Psi_j (x).}
Furthermore, the constraint $P_m P^m$ implies the linearized
equations of motion $\p_m\p^m \Psi_j =
\p_m\p^m \Psi_{\dot b} =0$.

So the physical states of the superparticle are described by a massless
$SO(8)$ vector $\Psi_j(x)$ and a massless
$SO(8)$ antichiral spinor $\Psi_{\dot a}(x)$ which are the physical
states of D=10 super-Yang-Mills theory. However, this description of
super-Yang-Mills theory only manifestly
preserves an $SO(8)$ subgroup of the super-Poincar\'e
group, and one would like a more covariant method for
quantizing the theory.
Covariant quantization can be extremely useful if one wants
to compute more than just
the physical spectrum in a flat background. For example, non-covariant  
methods are clumsy for computing scattering amplitudes or
for generalizing to curved backgrounds.

As will be shown in the following subsection, a manifestly
super-Poincar\'e covariant description of on-shell super-Yang-Mills
is possible using N=1 D=10 superspace. This covariant description will
later be obtained from quantization of a superparticle action involving
pure spinors.

\subsec{ Superspace description of super-Yang-Mills theory}

Although on-shell super-Yang-Mills theory can be described by the 
$SO(8)$ wavefunctions $\Psi_j(x)$
and $\Psi_{\dot a}(x)$ of \wfc\ satisfying the linearized
equations of motion $ \p_m\p^m \Psi_j =
\p_m\p^m \Psi_{\dot a} =0,$ there are more covariant descriptions 
of the theory. Of course, there is a Poincar\'e-covariant description
using an $SO(9,1)$ vector field $a_m(x)$ and an $SO(9,1)$
spinor field $\chi^\a(x)$ transforming in the adjoint representation
of the gauge group which satisfy the equations of motion 
\eqn\lorentz{
\p^m f_{mn} + ig [a^m,f_{mn}]=0,\quad  \g^m_{\a\b}(\p_m \chi^\b + ig
[a_m,\chi^\b])=0,}
and gauge invariance 
\eqn \gaui{\d a_m = \p_m s + ig [a_m,s],\quad
\d \chi^\a =  ig [\chi^\a,s], \quad \d f_{mn} = ig [f_{mn},s],}
where $f_{mn}= \p_{[m} a_{n]} + ig [a_m,a_n]$ is the Yang-Mills
field strength and $g$ is the super-Yang-Mills
coupling constant. 
However, there is also a super-Poincar\'e covariant
description 
using an $SO(9,1)$ spinor wavefunction $A_\a (x,\t)$ defined
in D=10 superspace. As will be explained below, on-shell super-Yang-Mills
theory
can be described by a spinor superfield 
$A_\a (x,\t)$ transforming in the adjoint representation
which satisfies the superspace equation
of motion\ref\supers{W. Siegel, {\it Superfields in Higher Dimensional
Spacetime}, Phys. Lett. B80 (1979) 220\semi E. Witten, {\it
Twistor-like Transform in Ten Dimensions}, Nucl. Phys. B266 (1986) 245.}
\eqn\eom{\g_{mnpqr}^{\a\b} (D_\a A_\b +i g A_\a A_\b) = 0}
for any five-form direction $mnpqr$,
with the gauge invariance
\eqn\superg{\d A_\a =  D_\a \Lambda + ig [A_\a,\Lambda]}
where $\Lambda(x,\t)$ is any scalar superfield and 
$$D_\a = {\p\over{\p\t^\a}} + {1\over 2}(\g^m\t)_\a \p_m$$
is the supersymmetric derivative.

One can also define field strengths constructed from $A_\a$ by
\eqn\fs{B_m = {1\over 8}\g_m^{\a\b} (D_\a A_\b +i g A_\a A_\b),\quad
W^\a = {1\over{10}}\g_m^{\a\b} (D_\a B^m - \p^m A_\a + ig [A_\a, B^m]),}
$$F_{mn} = \p_{[m} B_{n]} + ig [B_m, B_n] ={1\over 8} (\g_{mn})_\a{}^\b
(D_\b W^\a + ig \{A_\b, W^\a\}) $$
which transform
under the gauge transformation of \superg\ as 
\eqn\gauget{\d B_m = \p_m \Lambda + ig [B_m,\Lambda],\quad
\d W^\a = ig [W^\a,\Lambda],\quad \d F^{mn} = ig [F^{mn},\Lambda].}

To show that $A_\a(x,\t)$ describes on-shell super-Yang-Mills theory,
it will be useful to first note that in ten dimensions
any symmetric bispinor $f_{\a\b}$ can be decomposed in terms
of a vector and a five-form as
$f_{\a\b} = \g^m_{\a\b} f_m + \g^{mnpqr}_{\a\b} f_{mnpqr}$
and any antisymmetric bispinor $f_{\a\b}$ can be decomposed in terms
of a three-form as $f_{\a\b} = \g^{mnp}_{\a\b} f_{mnp}.$
Since $\{D_\a,D_\b\}= \g^m_{\a\b}\p_m$, one can check that 
$\d A_\a =  D_\a \Lambda + ig[A_\a,\Lambda]$ 
is indeed a gauge invariance of \eom.

Using $\Lambda(x,\t) = h_\a(x) \t^\a + j_{\a\b} (x)\t^\a \t^\b,$ one can
gauge away $(A_\a(x))|_{\t=0}$ and the
three-form part of $(D_\a A_\b(x))|_{\t=0}$. Furthermore, equation
\eom\ implies that the five-form part of 
$(D_\a A_\b(x))|_{\t=0}$ vanishes. So the lowest non-vanishing
component of $A_\a(x,\t)$ in this gauge is the vector component
$(D\g_m A(x))|_{\t=0}$ which will be defined as
$8 a_m(x)$. Continuing this type of argument to
higher order in $\t^\a$, one finds that there exists a gauge choice
such that
\eqn\compa{A_\a(x,\t) ={1\over 2}
(\g^m \t)_\a a_m(x) + {i\over{12}}(\t\g^{mnp}\t) (\g_{mnp})_{\a\b} 
\chi^\b(x) +  ... }
where $a_m(x)$ and $\chi^\b(x)$ are $SO(9,1)$ vector and spinor fields 
satisfying \lorentz\
and where the component fields in $...$ are functions of spacetime
derivatives of $a_m(x)$
and $\chi^\b(x)$.
Furthermore, this gauge choice leaves the residual gauge transformations
of \gaui\ where $s(x) = (\Lambda(x))|_{\t=0}$. Also,
one can check that the $\t=0$ components of the superfields
$B_m$, $W^\a$ and $F_{mn}$ of \fs\
are $a_m$, $\chi^\a$ and $f_{mn}$ respectively.
So the equations of motion and gauge invariances of \eom\ and
\superg\ correctly
describe on-shell super-Yang-Mills theory.

One would now like to obtain this super-Poincar\'e covariant description
of super-Yang-Mills
theory by quantizing the superparticle. 
Since the super-Yang-Mills spectrum contains a massless vector, one expects
the covariant superparticle constraints to generate the spacetime
gauge invariances of this vector. Note that these constraints
are not present in the gauge-fixed action of \lcaction\ since 
$\Psi_j$ describes only the transverse degrees of freedom
of the $SO(9,1)$ vector. Before describing the covariant
constraints which generate the gauge invariances of this vector,
it will be useful to first review the worldline action for Chern-Simons
theory which also has constraints related to spacetime gauge invariances.

\subsec{Worldline description of Chern-Simons theory}

Since the gauge invariance of a massless vector field
is $\d A_\mu = \p_\mu \Lambda$, one might guess that the
worldline action for such a 
field should contain the constraints $P_\mu.$
Although these constraints are too strong for describing Yang-Mills
theory, they are just right for describing D=3 Chern-Simons theory where
the field-strength of $A_\mu$ vanishes on-shell.

As was shown in \ref\cs{E. Witten, {\it Chern-Simons Gauge Theory
as a String Theory}, Prog. Math. 133 (1995) 637, hep-th/9207094.}, 
Chern-Simons theory can be described using the
worldline action\foot{Although \cs\ discusses only
a worldsheet action for Chern-Simons string theory, 
the methods easily generalize
to a worldline action.}
\eqn\csaction{
S=\int d\tau (\dot x^\mu P_\mu + l^\mu P_\mu)}
where $\mu=0$ to 2 and $l^\mu$ are Lagrange multipliers for the constraints.
Since the constraints are first-class, the action can be quantized
using the BRST method. After gauging $l^\mu=-\half P^\mu$, the gauge-fixed
action is 
\eqn\csgf{
S=\int d\tau (\dot x^\mu P_\mu -\half P^\mu P_\mu
+ \dot c^\mu b_\mu)}
with the BRST operator
\eqn\csbrst{Q = c^\mu P_\mu}
where $(c^\mu,b_\mu)$ are fermionic Fadeev-Popov ghosts and anti-ghosts.

To show that the cohomology of the BRST operator describes Chern-Simons
theory, note that the most general wavefunction
constructed from a ground state
annihilated by $b^\mu$ is 
\eqn\csfield{\Psi(c,x) = C(x) + c^\mu A_{\mu}(x) + {i\over 2}\e_{\mu\nu\rho}
c^\mu c^\nu A^{*\rho}(x) + {i\over 6}
\e_{\mu\nu\rho}c^\mu c^\nu c^\rho C^*(x)}
where the expansion in $c^\mu$ terminates since $c^\mu$ is fermionic.
One can check that 
\eqn\qphi{Q\Psi= -i c^\mu \p_\mu C -{i\over 2} c^\mu c^\nu \p_{[\mu} A_{\nu]}
+ {1\over 6} 
\e_{\mu\nu\rho}c^\mu c^\nu c^\rho \p_\sigma A^{*\sigma}(x).}
So $Q\Psi=0$ 
implies that
$A_\mu(x)$ satisfies the equations of motion $\p_{[\mu} A_{\nu]}=0$
which is the linearized
equation
of motion of the Chern-Simons field. Furthermore,
if one defines the gauge parameter
$\Omega(c,x) = i\Lambda(x) - c^\mu \omega_{\mu}(x) + ...$, the gauge
transformation $\delta\Psi=Q\Omega$ implies $\d A_\mu = \p_\mu\Lambda$
which is the linearized gauge transformation of the Chern-Simons field.

If one defines physical fields in BRST quantization to
carry ghost-number one, one finds that 
the spacetime ghosts carry ghost-number zero, the antifields carry
ghost number two, and the antighosts carry ghost-number three. From
the equations of motion and gauge invariances $Q\Psi=0$ and $\d\Psi=Q\O$, 
one learns that the gauge invariances of the antifields are related
to the equations of motion of the fields, and the equations of motion
of the ghosts are related to the gauge invariances of the fields.
For example, from $Q\Psi=0$ and $\d\Psi=Q\O$ for the Chern-Simons
wavefunction of \csfield, one learns that 
$A^{*\rho}$ satisfies
the equation of motion $\p_\sigma A^{*\sigma}=0$ with the gauge invariance
$\d A^{*\sigma} = \e^{\sigma\mu\nu}\p_{\mu} w_{\nu}$, which are the
linearized equations of motion and gauge invariance of the Chern-Simons
antifield. And the remaining fields, $C(x)$ and $C^*(x)$, describe
the spacetime ghost and antighost of Chern-Simons theory.

These equations of motion and gauge invariances can be obtained from
the Batalin-Vilkovisky version \ref\BV{I.A. Batalin and G.A.
Vilkovisky, {\it Quantization of Gauge Theories with Linearly
Dependent Generators}, Phys. Rev. D28 (1983) 2567.} of the
abelian Chern-Simons spacetime action
\eqn\css{{\cal S} = \int d^3 x (\half \e^{\mu\nu\rho} A_\mu \p_\nu A_\rho
+ i A^{*\mu} \p_\mu C),}
where, in addition to the usual Chern-Simons action for $A_\mu$,
there is a term coupling the antifield $A^{*\mu}$ to the gauge variation
of $A_\mu$. The action of \css\ can be written compactly in terms of the
wavefunction $\Psi$ of \csfield\ as 
\eqn\comp{{\cal S} = \half\int d^3 x \langle \Psi Q \Psi \rangle}
where $\langle ~~\rangle$ is normalized such that $\langle
c^\mu c^\nu c^\rho \rangle = i\e^{\mu\nu\rho}$.

Up to now, only abelian Chern-Simons theory has been discussed,
but it is easy to generalize to the non-abelian case. For example, the
Batalin-Vilkovisky version of the non-abelian Chern-Simons action is
\eqn\csna{{\cal S} = Tr \int d^3 x (\e^{\mu\nu\rho} (\half
A_\mu \p_\nu A_\rho + {i g\over 3} A_\mu A_\nu A_\rho) }
$$+
i A^{*\mu} (\p_\mu C + ig [A_\mu, C]) - g C C C^* ),$$
which can be written compactly as
\eqn\compna{{\cal S} = Tr\int d^3 x \langle \half\Psi Q \Psi
+{g\over 3} \Psi \Psi \Psi \rangle}
where $g$ is the Chern-Simons coupling constant and the fields in
$\Psi$ of \csfield\ now carry Lie algebra indices.
Note that the non-linear equations of motion and gauge invariances associated
with this action are
\eqn\nonlin{Q\Psi + g\Psi \Psi =0,\quad \d\Psi = Q\Omega +g [\Omega,\Psi].}
Using intuition learned from this worldline description of 
Chern-Simons theory, it will now be shown how to quantize the superparticle
in a similar manner. 

\subsec{ Pure spinor description of the superparticle}

In the case of Chern-Simons theory, the gauge transformation
$\d A_\mu = \p_\mu\Lambda$ was generated by the constraints $P_\mu$.
So for the superparticle, the gauge transformation $\d A_\a = D_\a\Lambda$
suggests using the constraints $d_\a$.
However, the constraints $d_\a$ are not
all first-class, so 
\eqn\brst{Q = \l^\alpha d_\alpha}
would not be a 
nilpotent operator for generic $\l^\a$. 
But since \antic\ implies that 
$Q^2= (\l^\a d_\a)^2 = -{i\over 2}\l^\alpha \l^\beta 
\gamma^m_{\alpha \beta} P_m$, $Q$ is nilpotent if
$\l^\alpha$ satisfies the pure spinor condition
\eqn\pure{ \l^\a \g^m_{\a\b} \l^\b =0}
for $m=0$ to 9. 
Note that $\l^\a$ must be complex in order to have solutions to
\pure. However, its complex conjugate $\bar\l^\a$ never appears
in the formalism so one is free to define $\l^\a$ to be a hermitian 
operator.
Defining 
$(\l^\a)^\dagger = \l^\a$ does not lead to any
inconsistencies since $\l^\a$ carries ghost number and therefore
does not have any $c$-number eigenvalues.
In other words, $\l^\a (\l^\b)^\dagger = \l^\a \l^\b$ takes states
of ghost-number $g$ to states of ghost-number $g+2$. So 
$\l^\a (\l^\a)^\dagger$ has no $c$-number eigenvalues and there
is therefore no reason that it should be positive-definite.

The pure spinor condition of \pure\ appears strange since bosonic
ghosts in the BRST formalism are normally unconstrained and come
from gauge-fixing fermionic Lagrange multipliers. However, as will
now be argued, the BRST operator and pure spinor constraint of
\brst\ and \pure\ can be derived by starting with the Brink-Schwarz
superparticle in semi-light-cone gauge, adding additional
fermionic degrees of freedom and gauge invariances, and then gauge-fixing
in a non-standard manner. 

The action of \lcaction\ for the Brink-Schwarz superparticle
in semi-light-cone gauge is 
\eqn\lcacttwo{
\int d\tau (\dot x^m P_m + {1\over 2}\dot S_a S_a + e P^m P_m) }
where $m=0$ to 9, $\a=1$ to 16, and $a=1$ to 8. Suppose one
now introduces a new set of $(p_\a,\t^\a)$ variables which are
unrelated to $S_a$ and defines $d_\a=p_\a +
\half P_m (\g^m\t)_\a$. Using $\{d_\a,d_\b\}= -i P_m\g^m_{\a\b}$
and $\{S_a,S_b\}=i\d_{ab}$, one can check that
\eqn\hatdefd{\hat d_\a = d_\a + (\g_m\g^+ S)_\a P^m 
(P^+)^{-\half}}
describes first-class constraints which close to
$\{\hat d_\a,\hat d_\b\} = -{i\over {2 P^+}} P_m P^m \g^+_{\a\b}$. So
\lcacttwo\ is equivalent to 
\eqn\lcactthree{ S=
\int d\tau (\dot x^m P_m + \dot \t^\a p_\a + 
{1\over 2}\dot S_a S_a + e P^m P_m + f^\a \hat d_\a) }
where $f^\a$ are fermionic Lagrange multipliers.
Since $\hat d_\a$ are first-class, they could be used to gauge
$\t^\a=0$ which would return \lcactthree\ to the original action
of \lcacttwo. 

Using the usual BRST method, the action of \lcactthree\ can
be gauge-fixed to 
\eqn\lcactfour{
S=\int d\tau (\dot x^m P_m -\half P^m P_m + \dot \t^\a p_\a +
{1\over 2}\dot S_a S_a 
+\dot c b +\dot{\hat\l}^\a 
\hat w_\a)}
together with the BRST operator
\eqn\hatbrst{\hat Q=\hat\l^\a \hat d_\a + c P^m P_m + {i\over{4 P^+}} b 
(\hat\l\g^+\hat\l)}
where $\hat\l^\a$ is an unconstrained bosonic spinor variable which comes
from gauge-fixing $f^\a=0$.
To relate $\hat Q$ with $Q=\l^\a d_\a$, it will first be argued
that the cohomology of $\hat Q$ is equivalent to the cohomology
of $Q' = \l'^\a\hat d_\a$ in a Hilbert space without
$(b,c)$ ghosts and where $\l'^\a$ is constrained to satisfy
$\l'\g^+\l'=0$ (but is not constrained to satisfy
$\l'\g^j\l'=0$ or $\l'\g^-\l'=0$). To show that $Q'$
has the same cohomology as $\hat Q$, consider a state $V$ annihilated
by $Q'$ up to terms proportional to $\l'\g^+\l'$, i.e.
$Q' V= (\l'\g^+\l')W$ for some $W$. Then
$(Q')^2=-{i\over{4P^+}}(\l'\g^+\l')P^m P_m $ implies that 
$Q'W= -{i\over{4P^+}} P^m P_m  V$. Using this information, one can
check that $\hat V= V + 4i P^+ c W$ is annihilated by $\hat Q$.
Furthermore, if $V$ is BRST-trivial up to terms involving
$\l'\g^+\l'$, i.e. $V=Q'\Omega + (\l'\g^+\l')Y$ for some $Y$,
then $V+4i P^+ 
c W= \hat Q(\Omega -4i P^+ cY)$, so $\hat V$ is also BRST-trivial.
So any state in the cohomology of $Q'$ is in the cohomology of
$\hat Q$, and reversing the previous arguments, one can show that
any state in the cohomology of $\hat Q$ is in the cohomology of
$Q'$. 

Finally, it will be shown that the cohomology of $Q'=\l'^\a \hat d_\a$
is equivalent to the cohomology of $Q=\l^\a d_\a$ where $\l^\a$ is
a pure spinor and the Hilbert space is independent of $S_a$.
Since $(\g^+\l')_{\dot a}$
is a null SO(8) antichiral spinor, it is preserved up to a phase
by some U(4) subgroup of SO(8). Under this U(4) subgroup, the chiral
SO(8) spinor $(\g^-\l')_a$ splits into a 4 and $\bar 4$ 
representation which will be called $(\g^-\l')_A$ and $(\g^-\l')_{\bar A}$
for $A,\bar A=1$ to 4. Similarly, the chiral SO(8) spinors
$(\g^+ d)_a$ and $S_a$ split into the representations
$[(\g^+ d)_A, (\g^+ d)_{\bar A}]$ and
$[S_A, S_{\bar A}]$.
Note that the $4$ and
$\bar 4$ representations are defined with respect to the null
spinor $(\g^+\l')_{\dot a}$ such that
$\s_j^{A \dot a}(\g^+\l')_{\dot a}$ is zero for $j=1$ to 8, and 
$\s_j^{\bar A \dot a}(\g^+\l')_{\dot a}$ is non-zero.
After performing a similarity transformation which shifts
$S_A \to S_A + (P^+)^{-\half} (\g^+ d)_A$, one finds that $Q'$ transforms as
\eqn\similbrst{Q' \to e^{-i S_{\bar A} (\g^+ d)_A (P^+)^{-\half} } 
Q' e^{i S_{\bar A}(\g^+ d)_A (P^+)^{-\half}} }
$$= (\g^+\l')_{\dot a} (\g^- d)_{\dot a} + (\g^-\l')_A (\g^+ d)_{\bar A}
+ (\g^-\l')_{\bar A} S_A \sqrt{P^+} .$$
So $Q' = \l^\a d_\a + (\g^-\l')_{\bar A} S_A \sqrt{P^+}$ 
where $\l^\a$ is a pure
spinor defined by
\eqn\lrelpure{[(\g^+\l)_{\dot a}, (\g^-\l)_A, (\g^-\l)_{\bar A}]=
[(\g^+\l')_{\dot a}, (\g^-\l')_A, 0].} Using the
standard quartet argument, the cohomology of $Q'=Q +(\g^-\l')_{\bar A}
S_A \sqrt{P^+}$ is equivalent
to the cohomology of $Q=\l^\a d_\a$ in the Hilbert space independent
of $(\g^-\l')_{\bar A}$, $S_A$, and its conjugate momenta 
$(\g^+ w')_{A}$ and $S_{\bar A}$.
So the Brink-Schwarz superparticle action has been shown to be
equivalent to the action
\eqn\bsequ{S=\int d\tau (\dot x^m P_m -\half P^m P_m + \dot \t^\a p_\a +
\dot\l^\a w_\a)}
together with the BRST operator $Q=\l^\a d_\a$ where $\l\g^m\l=0$.

Although the above derivation of the pure spinor description from
the Brink-Schwarz superparticle was not manifestly Lorentz covariant,
the final 
result of \bsequ\ is manifestly covariant. As will
be shown in the next subsection, quantization using this description
provides a manifestly super-Poincar\'e covariant description of
D=10 super-Yang-Mills theory. 

\subsec{Covariant quantization of the D=10 superparticle}

The most general super-Poincar\'e covariant wavefunction that can
be constructed from
$(x^m,\t^\a,\l^\a)$ is
\eqn\spf{\Psi(x,\t,\l) = C(x,\t) + \l^\a A_\a(x,\t) + 
(\l\g^{mnpqr}\l) A^*_{mnpqr}(x,\t) + \l^\a \l^\b \l^\g C^*_{\a\b\g}(x,\t) 
+ ...}
where $...$ includes superfields with more than three powers of $\l^\a$.
Note that
the names for the superfields appearing in \spf\ have been chosen to coincide
with the names for the Chern-Simons fields in \csfield. As in Chern-Simons,
the ghost-number zero superfield $C$ contains the spacetime ghost,
the ghost-number one superfield $A_\a$ contains the super-Yang-Mills
fields, the ghost-number two superfield $A^*_{mnpqr}$ contains
the super-Yang-Mills antifields, and the ghost-number three
superfield $C^*_{\a\b\g}$ contains the spacetime antighost. All
superfields in $...$ with ghost-number greater than three will have trivial
cohomology.

For example, 
$Q\Psi = -i\l^\a D_\a C -i \l^\a \l^\b D_\a A_\b + ...$,
so $Q\Psi=0$ implies that $A_\a(x,\t)$ satisfies the equation of motion
$\l^\a \l^\b D_\a A_\b=0$. But since 
$\l^\a \l^\b$ is proportional to $(\l\g^{mnpqr}\l) \g_{mnpqr}^{\a\b}$, 
this implies
that $D\g^{mnpqr}A=0$, which is the linearized version of the
super-Yang-Mills
equation of motion of \eom. Furthermore, if one defines the
gauge parameter $\Omega= i\Lambda + \l^\a \omega_\a + ...$, the
gauge transformation $\d\Psi = Q\Omega$ implies $\d A_\a = D_\a \Lambda$
which is the linearized super-Yang-Mills gauge transformation  
of \superg.

So as described in \compa, $A_\a(x,\t)$ contains 
the on-shell super-Yang-Mills gluon and
gluino, $a_m(x)$ and $\chi^\a(x)$, which satisfy the linearized equations
of motion and gauge invariances
$$\p^m \p_{[m} a_{n]} = \g^m_{\a\b} \p_m\chi^\b =0,\quad \d a_m = 
\p_m s.$$
And since gauge invariances of 
antifields correspond to equations of motion of fields, one
expects to have antifields
$a^{*m}(x)$ and $\chi^*_\a(x)$ in the cohomology of $Q$
which satisfy
the linearized equations of motion and
gauge invariances
\eqn\afeom{\p_m a^{*m}=0,\quad \d a^{*m} = \p_n (\p^n s^m - \p^m s^n),\quad
\d\chi_\a^* = \g^m_{\a\b}\p_m \kappa^\b}
where $s^m $ and $\kappa^\b$ are gauge parameters.
Indeed, these antifields $a^{*m}$ and $\chi^*_\a$ appear
in components of the
ghost-number $+2$ superfield $A^*_{mnpqr}$ of \spf.
Using $Q\Psi=0$ and $\d\Psi = Q\Omega$, $A^*_{mnpqr}$
satisfies the linearized
equation of motion $\l^\a(\l\g^{mnpqr}\l) D_\a A^*_{mnpqr}=0$
with the linearized
gauge invariance $\d A^*_{mnpqr} = \g_{mnpqr}^{\a\b} D_\a \omega_\b$.
Expanding $\omega_\a$ and $A^*_{mnpqr}$ in components, one learns
that 
$A^*_{mnpqr}$ can be gauged to the form
\eqn\astar{A^*_{mnpqr} = (\t\g_{[mnp}\t)(\t\g_{qr]})^\a \chi^*_\a(x) +
(\t\g_{[mnp}\t)(\t\g_{qr]s}\t) a^{*s}(x) + ...}
where $\chi^*_\a$ and $a^{*s}$ satisfy the equations of motion 
and residual gauge invariances of \afeom, and $...$ involves
terms higher order in $\t^\a$ which depend on derivatives of
$\chi^*_\a$ and $a^{*s}$.

In addition to these fields and antifields, one also expects
to find the Yang-Mills ghost $c(x)$ and antighost $c^*(x)$
in the cohomology of $Q$. The ghost $c(x)$ is found 
in the $\t=0$ component of the ghost-number zero
superfield, $C(x,\t) = c(x) + ...$, and the antighost
$c^*(x)$ is found in the $(\t)^5$ component of the
ghost-number $+3$ superfield,
$C^*_{\a\b\g}(x,\t)= ... + c^*(x) (\g^m\t)_\a (\g^n\t)_\b
(\g^p\t)_\g (\t\g_{mnp}\t) + ... .$ It was proven in \ref\superpart
{N. Berkovits, {\it Covariant Quantization of the Superparticle using       
Pure Spinors}, JHEP 0109 (2001) 016, hep-th/0105050.}
that the above
states are the only states in the cohomology of $Q$ and therefore,
although $\Psi$ of \spf\ contains superfields of arbitrarily high
ghost number, only superfields with ghost-number between zero and three
contain states in the cohomology of $Q$.

The linearized equations of motion and gauge invariances $Q\Psi=0$
and $\d\Psi =Q\Omega$ are easily generalized to the non-linear
equations of motion and gauge invariances
\eqn\nonls{Q\Psi + g\Psi \Psi =0,\quad \d\Psi = Q\Omega + g[\Psi,\Omega]}
where $\Psi$ and $\Omega$ transform in the adjoint representation of
the gauge group. For the superfield $A_\a(x,\t)$, \nonls\ implies
the super-Yang-Mills equations of motion and gauge transformations of
\eom\ and \superg.
Furthermore, the equation of motion and gauge transformation of \nonls\
can be obtained from the spacetime action\foot
{This spacetime action was first proposed to me by John Schwarz
and Edward Witten.}
\eqn\yma{{\cal S}= Tr\int d^{10}x \langle \half\Psi Q \Psi + 
{g\over 3}\Psi\Psi\Psi\rangle}
using the normalization definition that 
\eqn\norm{\langle (\l\g^m\t)(\l\g^n\t)(\l\g^p\t)(\t\g_{mnp}\t) \rangle =1.}
Although \norm\ may seem strange, it resembles the normalization of
\comp\ in that $\langle \Psi\rangle = c^*(x)$ where $c^*(x)$ is the
spacetime antighost. After expressing \yma\ in terms of component fields
and integrating out auxiliary fields, it should be possible to show
that \yma\ reduces to the standard Batalin-Vilovisky action for
super-Yang-Mills,
\eqn\bvym{{\cal S} = Tr\int d^{10}x ({1\over 4} f_{mn} f^{mn}
+\chi^\a \g^m_{\a\b} (\p_m\chi^\b +ig[a_m,\chi^\b])}
$$+ i a^{*m}(\p_m c + ig[a_m,c]) -g \chi^*_\a \{\chi^\a,c\} -g c c c^*).$$

Because the action of \yma\ only involves integration over
five $\t$'s, it is not manifestly spacetime supersymmetric. This
is not surprising since it is not known how to construct a
manifestly supersymmetric action for D=10 super-Yang-Mills. Nevertheless,
the equations of motion coming from this action have the same physical
content as the manifestly spacetime supersymmetric equations of motion
$Q\Psi + g\Psi\Psi=0$. This is because all components in
$Q\Psi + g\Psi\Psi=0$ with more than five $\t$'s are auxiliary equations
of motion. So removing these equations of motion only changes auxiliary
fields to gauge fields but does not affect the physical content of the
theory. By defining the normalization of \norm\ to involve $\l^\a(\sigma)$
and $\t^\a(\sigma)$ at the midpoint $\sigma={\pi\over 2}$ as in
\ref\bars{I. Bars, {\it Map of Witten's * to Moyal's *},
Phys. Lett. B517 (2001) 436, hep-th/0106157\semi I. Bars and Y. Matsuo, {\it
Computing in String Field Theory using the Moyal Star Product},
hep-th/0204260.}, it should be possible to generalize the action of 
\yma\ to a cubic open superstring field theory action.

\subsec{Pure spinor description for $d\neq 10$}

It is interesting to ask if the pure
spinor description can also be used to covariantly quantize the
superparticle when $d\neq 10$. Note that unlike the GS
superstring action, the Brink-Schwarz superparticle action is
invariant under $\k$-symmetry in any spacetime dimension.
If one defines a pure spinor in $d$ dimensions\foot{In 
arbitrary spacetime dimension,
this is not the pure spinor definition used by Cartan. For
example, in $d=11$, Cartan would define a pure spinor to satisfy
both $\l\g^m\l=0$ and $\l\g^{mn}\l=0$ \howetwo.} 
by $\l\g^m\l=0$ for $m=0$ to $d-1$,
a pure spinor contains $(3N-4)/4$ independent components where
$N$ is the number of components in an unconstrained $SO(d-1,1)$ spinor.
This counting can be derived using
a construction similar to the counting in $d=10$ where
$(\g^+\l)$ is a null $SO(d-2)$ spinor with $(N-2)/2$ components
and $(\g^-\l)$ is half of an $SO(d-2)$ spinor with $N/4$ components.
So $\l^\a$ has 2 components when $d=4$, 5 components when $5\leq d\leq 6$,
11 components when $7\leq d\leq 10$, and 23 components when $d=11$.

For $d=11$, it was shown in \ref\supermemb{N. Berkovits,
{\it Towards Covariant Quantization of the
Supermembrane},hep-th/0201151.} that the pure spinor
description correctly describes a superparticle whose physical spectrum
is linearized d=11 supergravity with 32 supersymmetries.
As discussed in \supermemb, physical states for the $d=11$ superparticle carry
ghost-number three and the state
$\Psi=\l^\a\l^\b\l^\g B_{\a\b\g}(x,\t)$ describes the $d=11$
supergravity multiplet where $B_{\a\b\g}$ is the spinor
component of the three-form superfield \ref\cederw{M. Cederwall, B.E.W.
Nilsson and D. Tsimpis, {\it Spinorial Cohomology and Maximally
Supersymmetric Theories}, JHEP 0202 (2002) 009, hep-th/0110069\semi
P.S. Howe, private communcation.}.
And for $7\leq d< 10$, one can easily check that the pure spinor
description correctly describes a superparticle whose physical
spectrum is a dimensional reduction of super-Yang-Mills with 16
supersymmetries. However, for $d\leq 6$, the situation is more
subtle. Note that a $d=6$ spinor is described by $\l_a^J$ where
$J=1$ to 2  is an SU(2) spinor index and $a=1$ to 4 is an
$SU^*(4)$ index. The constraint $\l\g^m\l=0$ implies 
$\l^J_a\l^K_b \e_{JK}=0$, which implies that $\l_a^J= c^J h_a$
for some $c^J$ and $h_a$. And for $d=4$, $\l\g^m\l=0$ implies
that either $\l_a=0$ or $\l_{\dot a}=0$ where $(a,\dot a)=1$ to 2
are the standard SU(2) Weyl indices.

Using techniques similar to the $d=10$ case,
one finds that for $5\leq d\leq 6$ or $d=4$, the
cohomology of $Q=\l^\a d_\a$ describes off-shell super-Yang-Mills
with 8 or 4 supersymmetries. As in $d=10$,
$Q\Psi=0$ implies that $\l^\a \l^\b D_\a A_\b=0$, which implies
that $D_{(\a} A_{\b)}=\g^m_{\a\b} B_m$ for some vector gauge field
$B_m$. However, unlike $d=10$, the theory is off-shell
since $D_{(\a} A_{\b)}= \g^m_{\a\b}
B_m$ does not impose equations of motion
when $d\leq 6$. This might seem surprising since the Brink-Schwarz
superparticle contains the $P_m P^m=0$ mass-shell 
constraint for any $d$. But note that for $d\leq 6$, there are also
subtleties in the light-cone quantization of the Brink-Schwarz
superparticle. When $d=6$, the light-cone $S_a$ variable
contains 4 components, which naively suggests $2^{4/2}=4$ states
in the physical spectrum instead of the 8 states of $d=6$ super-Yang-Mills.
And when $d=4$, $S_a$ contains 2 components, which naively suggests
$2^{2/2}=2$ physical states
instead of the 4 states of $d=4$ super-Yang-Mills.
Since light-cone quantization of the superparticle is not straightforward
in $d\leq 6$, it is not so surprising that there are
subtleties in the
pure spinor description in these dimensions.

\newsec {Covariant Quantization of the Superstring}

In this section, the pure spinor description of the superparticle
will be generalized to the superstring. Although there have been
several previous approaches to covariantly quantizing the superparticle,
this is the first approach which successfully generalizes to 
covariant quantization of the superstring. But before discussing
the pure spinor approach, it will be useful to discuss an alternative
approach of Siegel \siegelcs\
which contains some of the same features as the pure spinor
approach.

\subsec{Review of GS formalism using the approach of Siegel}

In conformal gauge, the classical
covariant GS action for the heterotic superstring is\ref\GS{M.B. Green and
J.H. Schwarz, {\it Covariant Description of Superstrings}, Phys. Lett. 
B136 (1984) 367.}
\eqn\het
{S_{het}=\int d^2 z [\half\Pi^m \bar\Pi_m +{1\over 4} 
\Pi_m \t^\a \g^m_{\a\b} \bar\p\t^\b
-{1\over 4}\bar\Pi_m \t^\a\g^m_{\a\b} \p\t^\b] + S_R}
where $x^m$ and $\t^\a$ are the worldsheet variables ($m=0$ to 9,
$\a=1$ to 16), $S_R$ describes the right-moving degrees of
freedom for the 
$E_8\times E_8$ or $SO(32)$ lattice, and
$\Pi^m = \p x^m + \half\t^\a \g^m_{\a\b} \p\t^\b$ and
$\bar\Pi^m = \bar\p x^m +\half \t^\a \g^m_{\a\b} \bar\p\t^\b$
are supersymmetric combinations of the momentum.
In what follows, the right-moving degrees of freedom play no
role and will be ignored. Also, all of the following remarks
are easily generalized to the Type I and Type II superstrings.

Since the action of \het\ is in conformal gauge, it needs to be
supplemented with the Virasoro constraint $T= -\half \Pi^m \Pi_m=0$.
Also, since the canonical momentum to $\t^\a$ does not appear in the
action, one has the Dirac constraint 
$p_\a = {\d\L}/{\d \p_0\t^\a}=\half(\Pi_m -{1\over 4}\t\g_m\p_1\t) (\g^m\t)_\a$
where $p_\a$ is the canonical momentum to $\t^\a$. 
If one defines
\eqn\defdone{
d_\a = p_\a -\half(\Pi_m -{1\over 4}\t \g_m\p_1\t ) (\g^m\t)_\a,}
one can use the canonical commutation relations to find
$\{d_\a, d_\b\}= i\g^m_{\a\b} \Pi_m,$
which implies (since $\Pi^m\Pi_m=0$ is a constraint)
that the sixteen Dirac constraints $d_\a$ have eight
first-class components and eight second-class components.
Since the anti-commutator of the second-class constraints is
non-trivial (i.e. the
anti-commutator 
is an operator $\Pi^+$ rather than a constant), standard Dirac quantization
cannot be used since it would involve inverting an operator.
So except in light-cone gauge (where the commutator becomes a constant),
the covariant Green-Schwarz formalism cannot be easily quantized.

In 1986, Siegel suggested an alternative approach in which
the canonical momentum to $\t^\a$ is an independent variable
using the free-field action \siegelcs
\eqn\siegel
{S=\int d^2 z [\half \p x^m \bar\p x_m + p_\a\bar\p\t^\a].}
In this approach, Siegel attempted to replace 
the problematic constraints of the covariant GS action with some suitable
set of first-class constraints constructed out of the 
supersymmetric objects $\Pi^m$, $d_\a$ and $\p\t^\a$ where
\eqn\defdtwo{d_\a=p_\a - \half(\p x^m +{1\over 4}\t\g^m\p\t) (\g_m\t)_\a}
is defined as in \defdone\ 
and is no longer constrained to vanish.
The first-class constraints should include the Virasoro constraint
$A=-\half \Pi^m \Pi_m
- d_\a \p\t^\a = -\half \p x^m \p x_m - p_\a \p\t^\a$
and the $\k$-symmetry generator 
$B^\a = \Pi^m (\g_m d)^\a.$ To get to light-cone gauge, one also needs
constraints such as $C^{mnp}= d_\a (\g^{mnp})^{\a\b} d_\b$ which is
supposed to replace the second-class constraints in $d_\a$. Although
this approach was successfully used for quantizing the superparticle
\ref\ilk{F. Essler, M. Hatsuda, E. Laenen, W. Siegel and J.
Yamron, {\it Covariant Quantization of the First Ilk Superparticle},
Nucl. Phys. B364 (1991) 67.},
a set of constraints which closes at the quantum level and which
reproduces the correct physical superstring spectrum was never found.

Nevertheless, the approach of Siegel has the advantage that
all worldsheet fields are free which makes it trivial to compute
the OPE's that 
\eqn\opeone{x^m(y) x^n(z)\to -2\eta^{mn} \log|y-z|,\quad
p_\a(y)\t^\b(z) \to \d_\a^\b (y-z)^{-1},}
\eqn\oped{d_\a(y)d_\b(z)\to
-{1\over{(y-z)}}\g^m_{\a\b}\Pi_m(z),\quad
d_\a(y)\Pi^m(z)\to
{1\over{(y-z)}}\g^m_{\a\b}\p\t^\b(z).}
This gives some useful clues about the appropriate ghost
degrees of freedom. Since $(\t^\a,p_\a)$
contributes $-32$ to the conformal anomaly, the total
matter contribution is $-22$ which is expected to be cancelled by
a ghost contribution of $+22$.
Furthermore, the spin contribution to the $SO(9,1)$ Lorentz currents 
in Siegel's approach is $M_{mn}=\half p\g_{mn}\t$, as compared with
the spin contribution to the $SO(9,1)$ Lorentz currents
in the RNS formalism which is $\psi_m\psi_n$.
These two Lorentz currents satisfy similar OPE's except for the 
numerator in the double pole of $M_{mn}$ with $M_{mn}$, which
is $+4$ in Siegel's approach and $+1$ in the RNS formalism.
This suggests that the worldsheet ghosts should have Lorentz currents
which contribute $-3$ to the double pole.

\subsec{Superstring quantization using pure spinors}

In fact, there exists an $SO(9,1)$ irreducible representation contributing
$c=22$ and with a $-3$ coefficient
in the double pole of its Lorentz current \cov.
This representation consists of a bosonic pure spinor $\l^\a$ satisfying the 
condition that 
\eqn\pure{\l^\a \g^m_{\a\b} \l^\b =0}
for $m=0$ to 9. To show that this representation has the desired
properties, it is useful to temporarily break manifest
Lorentz invariance by explicitly solving the pure spinor constraint
of \pure.

A parameterization of $\l^\a$
which preserves a $U(5)$ subgroup of (Wick-rotated)
$SO(10)$ is \cov\rnsmap
\eqn\paramone{\l^+ = e^s ,\quad \l_{ab} =  u_{ab},\quad
\l^a = -{1\over 8} e^{-s}\e^{abcde} u_{bc} u_{de}}       
where $a=1$ to 5,
$u_{ab}= - u_{ba}$ are ten independent variables,
and the SO(10)
spinor $\l^\a$ has been written in terms of its irreducible $U(5)$
components which transform as $(1_{{5\over 2}}, \overline{10}_{\half},
5_{-{3\over 2}})$ representations of $SU(5)_{U(1)}$. A simple
way to obtain these $U(5)$ representations is to write an $SO(10)$
spinor using $[\pm\pm\pm\pm\pm]$ notation
where Weyl/anti-Weyl spinors have an odd/even number of $+$ signs.
The $1_{5\over 2}$ component of $\l^\a$ is the component with five
$+$ signs, the
$\overline{10}_{\half}$ component has three $+$ signs, and the
$5_{-{3\over 2}}$ component has one $+$ sign.
The $\l^\a$
parameterization of \paramone\ is possible whenever $\l^+\neq 0$.

Using the above parameterization of $\l^\a$, one can define
the action $S_\l$ for the worldsheet ghosts as 
\eqn\actone{S_\l = \int d^2 z [\overline\p t \p s  - \half
v^{ab}\overline\p u_{ab}]}
where $t$ and $v^{ab}$ are the conjugate momenta to $s$ and $u_{ab}$
satisfying the OPE's
\eqn\opel{t(y)~ s(z) \to \log (y-z),\quad v^{ab} (y) ~u_{cd}(z) \to
\d_c^{[a} \d_d^{b]} (y-z)^{-1}.}
Note that the factor of $\half$ in the $v^{ab}
\overline\p u_{ab}$ term has been
introduced to cancel the factor of 2 from $u_{ab}=
-u_{ba}$.
Also note that $s$ and $t$ are chiral bosons, so their
contribution to \actone\ needs to be supplemented by a chirality constraint.

One can construct $SO(10)$
Lorentz currents $N^{mn}$ out of these free
variables as
\eqn\lorentz{N = {1\over{\sqrt{5}}}( {1\over 4}
u_{ab} v^{ab} + {5\over 2}\p t - {5\over 2}\p s), \quad
N_a^b =  u_{ac} v^{bc} -{1\over 5}\d_a^b u_{cd}v^{cd}, }
$$N^{ab} = e^s v^{ab}, \quad
N_{ab} =  e^{-s}(2 \p u_{ab} - u_{ab}\p t -2 u_{ab}\p s
+u_{ac} u_{bd} v^{cd} -\half
u_{ab} u_{cd} v^{cd}) $$
where $N^{mn}$ has been written in terms of its $U(5)$ components
$(N,N_a^b,N^{ab},N_{ab})$
which transform as $(1_0,24_0,10_2,\overline{10}_{-2})$ representations
of $SU(5)_{U(1)}$.
The Lorentz currents of \lorentz\ can be checked to satisfy the OPE's
\eqn\Nlope{N^{mn}(y) \l^\a (z) \to
\half(\g^{mn})^\a{}_\b {{\l^\b(z)}\over{y-z}},}
\eqn\nope{N^{kl}(y) N^{mn}(z) \to
{{\eta^{m[l} N^{k]n}(z) -
\eta^{n[l} N^{k]m}(z) }\over {y-z}} - 3
{{\eta^{kn} \eta^{lm} -
\eta^{km} \eta^{ln}}\over{(y-z)^2}}  .}
So although $S_\l$ of \actone\ is not manifestly Lorentz covariant, any OPE's
of $\l^\a$ and $N^{mn}$ which are
computed using this action are manifestly covariant.

In terms of the free fields, the stress tensor is
\eqn\stress{T_\l =  \half v^{ab}\p u_{ab}
+ \p t\p s + \p^2 s}
where the $\p^2 s$ term is included so that the Lorentz currents of
\lorentz\ are primary fields. This stress tensor has central charge $+22$
and can be written in manifestly Lorentz invariant notation as\ref\nbersh
{N. Berkovits and M. Bershadsky, unpublished.}
\eqn\covstress{T_\l =
  {1\over {10} } N_{mn} N^{mn}
- {1\over 8} J^2 - \p J}
where $J$ is defined in terms of the free fields
by
\eqn\hdef{J = {1\over 2} u_{ab} v^{ab} + \p t + 3 \p s.}
Note that
$J$ has no singularities with $N^{mn}$ and satisfies the OPE's
$$J(y) J(z) \to -4 (y-z)^{-2},\quad
J(y) \l^\a (z) \to (y-z)^{-1} \l^\a(z).$$
The operator $\oint J$ can be identified with the
ghost-number operator so that $\l^\a$ carries ghost number $+1$.

\subsec{ Physical vertex operators}

Physical states in the pure spinor formalism for the open 
superstring are defined as ghost-number one states in
the cohomology of $Q=\int \l^\a d_\a$ where
$\l^\a$ is constrained to satisfy
$\l\g^m\l=0$.
The constraint $\l\g^m\l=0$ implies that the
canonical momentum for $\l^\a$, which will be called
$w_\a$,
only appears in combinations which are invariant under the
gauge transformation 
\eqn\defwgauge{\d w_\a= (\g^m \l)_\a \L_m}
for
arbitrary
$\L_m$. This implies that $w_\a$ only appears in the
Lorentz-covariant
combinations $N_{mn}=\half :w\g_{mn}\l:$ and $J= :w_\a
\l^\a:$ where the normal-ordered expressions can be explicitly
defined using the parameterization of \lorentz\ and \hdef.
When $(mass)^2=
n/2$, open superstring vertex operators are constructed
from arbitrary combinations of $[x^m,\t^\a,d_\a, \l^\a,
N^{mn},J]$
which carry ghost number one and conformal weight $n$ at
zero     
momentum. Note that
$[d_\a,N_{mn},J]$ carry conformal weight one and $\l^\a$
carries
ghost number one. 

For example, the most general
vertex operator at $(mass)^2=0$
is 
\eqn\unintv{U=\l^\a A_\a(x,\t)}
where $A_\a(x,\t)$ is an
unconstrained
spinor superfield.
As was shown in subsection (2.5), $QU=0$ and $\d U=Q\Omega$ implies
$\g_{mnpqr}^{\a\b}D_\a A_\b=0$ and $\d A_\a = D_\a \Omega$,
which are the super-Maxwell equations of motion and gauge
invariances
written in terms of a spinor superfield.

At the next mass level, the physical 
states of the open superstring
form a massive spin-2 multiplet containing 128 bosons and 
128 fermions. Although it was not previously known how to
covariantly describe this multiplet in D=10 superspace, such a
superspace description was found
with Osvaldo Chand\'{\i}a using the pure spinor approach \ref\massch
{N. Berkovits and O. Chand\'{\i}a, {\it Massive Superstring Vertex Operator in
D=10 Superspace}, hep-th/0204121.}.
When $(mass)^2=\half$, the most general
vertex operator is
\eqn\VV{U=\p\l^\a A_\a(x,\t)+:\p\t^\b\l^\a B_{\a\b}(x,\t):
+
:d_\b \l^\a {C^\b}_\a(x,\t):}
$$+:\Pi^m \l^\a H_{m\a}(x,\t):+:J \l^\a E_\a(x,\t):
+
:N^{mn} \l^\a F_{\a mn}(x,\t):$$
where $:O^A \l^\a\Phi_{\a A}(x,\t)(z):$
=$\oint {dy\over{y-z}} O^A(y)~\l^\a(z)
\Phi_{\a A}(z)$ and
$\Phi_{\a A}(x,\t)$ are the various superfields appearing
in \VV. Using the OPE's of \oped, it was shown in \massch\ 
that $QU=0$ implies the equations
\eqn\eqs{(\g_{mnpqr})^{\a\b}[D_\a
B_{\b\g}-\g^s_{\a\g}H_{s\b}]=0,}
$$
(\g_{mnpqr})^{\a\b}[D_\a
H_{s\b}-\g_{s\a\g}{C^\g}_\b]=0,$$
$$
(\g_{mnpqr})^{\a\b}[D_\a{C^\g}_\b+ \d^\g_\a E_\b +\half
(\g^{st})^\g{}_\a F_{\b st}]=0,$$
$$
(\g_{mnpqr})^{\a\b}[D_\a
A_\b+B_{\a\b}+2\g^s_{\b\g}\p_s{C^\g}_\a-D_\b E_\a +
{1\over 2}
(\g^{st} D)_\b F_{\a st} ]$$
$$=
4\g_{mnpqr}^{\a\b} \g^{vwxys}_{\a\b} \eta_{st} K^t_{vwxy},$$
$$
(\g_{mnp})^{\a\b}[D_\a
A_\b+B_{\a\b}+2\g^s_{\b\g}\p_s{C^\g}_\a-D_\b E_\a +
{1\over 2}
(\g^{st} D)_\b F_{\a st}]$$
$$=
32\g_{mnp}^{\a\b}
\g^{wxy}_{\a\b} K^s_{wxys},$$
$$\g_{mnpqr}^{\a\b} D_\a E_\b =
\g_{mnpqr}^{\a\b} (\g^{vwxy}\g_s)_{\a\b} K^s_{vwxy},$$
$$\g_{mnpqr}^{\a\b} D_\a F_{\b}^{st}=
-\g_{mnpqr}^{\a\b} (\g^{vwxy}\g^{[s})_{\a\b}
K^{t]}_{vwxy},$$
where $K^s_{vwxy}$ is an arbitrary superfield.
And the gauge invariance $\d U= Q\O$ implies the gauge transformations
\eqn\gtransf{\d
A_\a=\O_{1\a}+2\g^m_{\a\b}\p_m\O^\b_2-
D_\a\O_4-{1\over 2}{(\g^{mn})^\b}_\a
D_\b\O_{5mn},}
$$\d B_{\a\b}=-D_\a\O_{1\b}+\g^m_{\a\b}\O_{3m},$$
$$\d
{C^\b}_\a=-D_\a\O^\b_2-\d^\b_\a\O_4-\half{(\g^{mn})^\b}_\a\O_{5mn},$$
$$\d H_{m\a}=D_\a\O_{3m}-\g_{m\a\b}\O^\b_2,$$
$$\d E_\a=D_\a \O_4,$$
$$\d F_{\a mn}=D_\a\O_{5mn},$$
where 
\eqn\geng{
\O=  :\p\t^\a\O_{1\a}(x,\t): + :d_\a\O^\a_2(x,\t):
+ :\Pi^m\O_{3m}(x,\t):}
$$+ :J\O_4(x,\t): + :N^{mn}\O_{5mn}(x,\t):,$$
and $:O^A \O_A(x,\t):$
=$\oint {dy\over{y-z}} O^A(y)~ \O_A(z).$
Using d=10 superspace techniques, it was argued in \massch\ that
the equations of motion and gauge transformations of \eqs\ and
\gtransf\ imply that the superfields $\Phi_{\a A}(x,\t)$ in \VV\
correctly
describe a
massive
spin-two multiplet with $(mass)^2=\half$.

To compute scattering amplitudes, one also needs vertex operators
in integrated form, $\int dz V$, where $V$ is 
usually obtained from the unintegrated
vertex operator $U$ by anti-commuting with the $b$ ghost. But since
there is no natural candidate for the $b$ ghost in
this formalism, 
one needs to use an alternative method for obtaining $V$ which is
from the relation
$[Q,V] = \p U$ \ref\bp
{N. Berkovits, M. Hatsuda and W. Siegel, {\it
The Big Picture}, Nucl. Phys. B371 (1992) 434, hep-th/9108021.}.
Using this alternative method, one finds for the open
superstring massless vertex operator that \cov 
\eqn\intv{
V= \p\t^\a A_\a(x,\t) +\Pi^m B_m(x,\t) +d_\a  W^\a(x,\t) +\half N_{mn}
F^{mn}(x,\t) .}
To show that $QV=\p U$, note that
\eqn\qv{QV = \p(\l^\a A_\a) + \l^\a \p\t^\b (-D_\a A_\b - D_\b A_\a
+\g_{\a\b}^m B_m)}
$$ + \l^\a \Pi^m (D_\a B_m - \p_m A_\a -\g_{m\a\b}W^\b)
+ \l^\a d_\b (-D_\a W^\b +{1\over 4} (\g^{mn})_\a{}^\b F_{mn})
+ \half\l^\a N_{mn} D_\a F^{mn}.$$
So $QV=\p U$ if the superfields satisfy
\eqn\qvsat{-D_\a A_\b - D_\b A_\a +\g_{\a\b}^m B_m=0,}
$$D_\a B_m - \p_m A_\a -\g_{m\a\b}W^\b=0,$$
$$-D_\a W^\b +{1\over 4} (\g^{mn})_\a{}^\b F_{mn}=0,$$
$$\l^\a \l^\b  (\g_{mn})_\b{}^\g D_\a F^{mn}=0,$$
which imply the super-Maxwell equations of subsection (2.2).
Note that the fourth equation of \qvsat\
is implied by the third equation since $\l^\a \l^\b D_\a D_\b W^\g =
\half(\l\g^m \l) \p_m W^\g =0$.
It is useful to note that in components,
\eqn\rns{
V = a_m (x) \p x^m +\half
 \p_{[m} a_{n]}(x) M^{mn} + \xi^\a(x) q_\a + O(\t^2),}
where $M^{mn}=
\half p\g^{mn}\t + N^{mn}$ is the spin contribution to the Lorentz current
and
$q_\a= p_\a +\half(\p x^m +{1\over{12}}\t\g^m\p\t) (\g_m\t)_\a$
is the spacetime-supersymmetry current. So
\rns\ closely resembles the RNS vertex operator \fms\
for the gluon
and gluino. If one drops the $\half N_{mn} F^{mn} $ term, the vertex operator
of \intv\ was suggested 
by Siegel \siegelcs\ based on superspace arguments. 

For the Type II superstring, the unintegrated massless vertex operator is
$U=\l^\a \hat\l^{\bh} A_{\a\bh}(x,\t,\hat\t)$ where
$\hat\l^\ah$ and $\hat\t^{\ah}$ are right-moving worldsheet fields
and the chirality of the
$\hat\a$ index depends if the superstring is IIA or IIB.
The physical state condition $QU=\hat QU =0$ and gauge invariance
$\d U = Q\hat\Omega + \hat Q\Omega$ 
where $\hat Q\hat\Omega=Q\Omega=0$
implies that 
\eqn\eomtwo{\g_{mnpqr}^{\a\b} D_\a A_{\b\gh} =  
\g_{mnpqr}^{\ah\gh} \hat D_\ah A_{\b\gh} = 
0, \quad \d A_{\a\bh} = D_\a \hat\Omega_\bh + \hat D_\bh \Omega_\a,}
$$\g_{mnpqr}^{\a\b} D_\a \Omega_\b =  
\g_{mnpqr}^{\ah\gh}\hat D_\ah \hat\Omega_\gh = 0$$
for any five-form direction
$mnpqr$, which are the linearized equations of motion
and gauge invariances 
of the Type IIA or Type IIB supergravity multiplet. The integrated form
of the closed superstring massless vertex operator 
is the left-right product of
the open superstring vertex operator of \intv\ and is given by
\eqn\closedvertex{
V_{SG} = \int d^2 z}
$$ [
\p\t^\a \bar\p\hht^\bh A_{\a \bh}(x,\t,\hht) +
\p\hat\t^\a \bar\Pi^m A_{\a m}(x,\t,\hht) +
\Pi^m \bar\p\hht^\ah A_{m \ah}(x,\t,\hht)
+\Pi^m \bar\Pi^n A_{m n}(x,\t,\hat\t)$$
$$+ d_\a (\bar\p\hht^\bh E^\a_\bh(x,\t,\hht) + \bar\Pi^m E^\a_m(x,\t,\hht))
+ \hd_\ah (\p\t^\b E^\ah_\b (x,\t,\hht)+ \Pi^m E^\ah_m(x,\t,\hht)) $$
$$ +\half N_{mn}
 (\bar\p\hht^\bh \O^{mn}_\bh(x,\t,\hht) + \bar\Pi^p\O^{mn}_p(x,\t,\hht))
+\half \hat N_{mn}
(\p\t^\b \hat\O^{mn}_\b (x,\t,\hht)+ \Pi^p \hat\O^{mn}_p(x,\t,\hht)) $$
$$+ d_\a \hd_\bh P^{\a\bh}(x,\t,\hht) + N_{mn} \hd_\ah C^{mn\ah}(x,\t,\hht)
+ d_\a \hat N_{mn} \hat C^{\a mn} (x,\t,\hht)
+ N_{mn} \hat N_{pq} S^{mnpq}(x,\t,\hht)].$$

\subsec { Tree-level scattering amplitudes}

As usual, the $N$-point tree-level open superstring
scattering amplitude will be defined as
the correlation function 
of $3$ unintegrated vertex operators $U_r$
and $N-3$ integrated vertex operators $\int dz V_r$ as
\eqn\amplcor{A=\int dz_4 ...\int dz_N ~\langle U_1(z_1) U_2(z_2) U_3(z_3)
\prod_{r=4}^N V_r(z_r)\rangle.}
For massless external states, the vertex operators are given in \unintv\ 
and \intv. 

The first step to evaluate the correlation function
is to eliminate all worldsheet fields of non-zero dimension
(i.e. $\p x^m$, $\p\t^\a$, $p_\a$, $J$ and $N^{mn}$)
by using their OPE's with other worldsheet fields and 
the fact that they vanish at $z\to\infty$.
One then integrates over the $x^m$ zero modes
to get a Koba-Nielsen type formula, 
\eqn\koba{A = \int dz_4 ... dz_N  \langle \l^\a \l^\b \l^\g
f_{\a\b\g}(z_r, k_r, \eta_r, \t) \rangle}
where $\l^\a\l^\b\l^\g$ comes from the three unintegrated
vertex operators and $f_{\a\b\g}$ is some function of
the $z_r$'s, the momenta $k_r$, the polarizations 
$\eta_r$, and the remaining $\t$ zero modes.

One would like to define the correlation function 
$\langle \l^\a \l^\b \l^\g f_{\a\b\g}\rangle$
such
that $A$ is supersymmetric and gauge invariant. An obvious way to
make $A$ supersymmetric is to require that the correlation function
vanishes unless all sixteen $\t$ zero modes are present, but this
gives the wrong answer by dimensional analysis. The correct answer
comes from realizing that $Y= \l^\a\l^\b\l^\g f_{\a\b\g}$
satisfies the constraint $QY=0$ when the external states are on-shell.
Furthermore, gauge invariance implies that $\langle Y\rangle$ should
vanish whenever $Y=Q\Omega$. 

As discussed in subsection (2.5),
there is precisely one
state 
in the cohomology of $Q$ 
at zero momentum and ghost-number three
which is $(\l\g^m\t)(\l\g^n\t)(\l\g^p\t)
(\t\g_{mnp}\t)$. So if 
\eqn\expa{f_{\a\b\g}(\t) = A_{\a\b\g}+ \t^\d B_{\a\b\g\d} + ... +
(\g^m\t)_\a (\g^n\t)_\b (\g^p\t)_\g (\t \g_{mnp}\t) F + ...,}
it is natural to define 
\eqn\harm{\langle \l^\a \l^\b \l^\g f_{\a\b\g}(z_r,k_r,\eta_r,\t)\rangle
 = F(z_r,k_r,\eta_r) .}
This definition is supersymmetric when all external states
are on-shell since 
$$(\l\g^m\t)(\l\g^n\t)(\l\g^p\t)(\t \g_{mnp}\t)$$ 
cannot be written as the supersymmetric variation of a quantity
which is 
annihilated
by $Q$. And the definition is gauge invariant since 
$$(\l\g^m\t)(\l\g^n\t)(\l\g^p\t)(\t \g_{mnp}\t)\neq Q\Omega$$
for any $\Omega$.
Note that
\harm\ can be interpreted as integration over an on-shell harmonic
superspace involving five $\t$'s since
$\langle 
\l^\a\l^\b\l^\g f_{\a\b\g}\rangle$ is proportional to
\eqn\proppr{({\p\over{\p\t}} \g_m)^\a ({\p\over{\p\t}} \g_n)^\b
({\p\over{\p\t}} \g_p)^\g ({\p\over{\p\t}}\g^{mnp}{\p\over{\p\t}}) 
f_{\a\b\g}|_{\t=0} =
\int (d^5\t)^{\a\b\g} f_{\a\b\g}.}

For three-point scattering, $A=\langle \l^\a A_\a^1(z_1)~ \l^\b A_\b^2 (z_2)
~\l^\g A_\g^3(z_3)
\rangle$, it is easy to check that the prescription of \harm\ reproduces
the usual super-Yang-Mills cubic vertex. In the gauge of \compa, each $A_\a$
contributes one, two or three $\t$'s. If the five $\t$'s are distributed
as $(1,1,3)$, one gets the $a_m^1 a_n^2 \p^{[m} a^{3 n]}$ vertex for
three gluons,
whereas if they are distributed as $(2,2,1)$, one gets the 
$(\xi^1 \g^m\xi^2) a_m^3$ vertex for two gluinos and one gluon. 
Together with Brenno Vallilo,
it was proven that the above prescription agrees with the standard RNS
prescription of \fms\
for N-point massless tree amplitudes involving up to four
fermions \ampval. 
The relation of \rns\ to the RNS massless vertex operator was
used in this proof, and the restriction on the number of fermions comes
from the need for different pictures in the RNS prescription.
Furthermore, using the map from on-shell states in the pure spinor BRST
cohomology to on-shell states in the RNS formalism, it was argued in 
\rnsmap\
that tree amplitudes involving massive states must also agree
with the RNS prescription.

\newsec{Quantization of the Superstring in a Curved Background}

Although it is not known how to covariantly quantize the GS
superstring, one can construct the classical GS superstring
action in a curved background. It has been shown that when the
background fields satisfy their on-shell equations to lowest order
in $\a'$, the classical worldsheet action is invariant under 
$\k$-symmetry. However, because of quantization problems, it is
not known how to compute $\a'$ corrections to the background
equations of motion using the GS formalism.

As will be reviewed here, one can
use the pure spinor description to construct an analogous
action for the superstring in
a curved background. In this case, classical BRST invariance will
imply the on-shell equations for the background to lowest order
in
$\a'$. Since quantization is straightforward using the pure spinor
description, one can now compute $\a'$ corrections to the background
equations by requiring quantum BRST invariance of the action. 
Note that in the pure spinor description, 
the equations coming from classical BRST invariance 
are expected
to imply that the action is conformally invariant to one-loop
order. Since the one-loop beta function vanishes, 
it is sensible to ask if there are finite
corrections to the background equations coming from one-loop BRST invariance.
Similarly, $n$-loop BRST invariance is expected to imply
$(n+1)$-loop conformal invariance of the action, so this method
can in principle be extended to all orders in $\a'$.

\subsec{Relation between $\k$-symmetry and classical BRST invariance}

The fact that classical BRST invariance in the pure spinor description
is related to $\k$-symmetry in the GS description can be understood
by computing the Poisson brackets of $Q=\int \l^\a d_\a$ with
the worldsheet fields. One finds that
\eqn\brstx{\d_Q x^m=\l\g^m\t,\quad \d_Q\t^\a=\l^\a,\quad
\d_Q d_\a=- \Pi^m(\g_m\l)^\a,\quad
\d_Q w_\a=d_\a,}
which resemble the $\k$-symmetry transformations
\eqn\kappatr{\d x^m=\xi\g^m\t,\quad \d\t^\a=\xi^\a,}
where $\xi^\a = \Pi^m (\g_m\k)^\a$. As shown by Oda and Tonin \ref\oda
{I. Oda and M. Tonin, {\it On the Berkovits Covariant Quantization
of GS Superstring}, Phys. Lett. B520 (2001) 398, hep-th/0109051.},
this relation is useful for constructing BRST-invariant actions
from $\k$-invariant GS actions.

If the GS action $S_{GS}$ satisfies
$\d S_{GS}=0$ under \kappatr\ up to the
Virasoro constraint $\Pi_m\Pi^m=0$ when 
$\xi^\a=\Pi^m (\g_m\k)^\a$, then when $\xi^\a$ is
arbitrary, $\d S_{GS}=\int d^2 z \Pi_m (\xi\g^m\O)$
for some $\O^\a$. Since $S_{GS}$ is independent 
of $d_\a$ and $w_\a$, this implies from \brstx\ that the BRST transformation
of $S_{GS}$ is $\d_Q S_{GS}=  \int d^2 z \Pi^m (\l\g^m\O)$.
One can therefore define a classically BRST-invariant action as
\eqn\clbrst{S_{BRST}= S_{GS} +\int d^2 z \d_Q (w_\a \O^\a)}
$$= S_{GS} + \int d^2 z [ d_\a \O^\a + w_\a \d_Q\O^\a ].$$
Although $Q^2=0$ naively implies that $\int d^2 z \d_Q (w_\a \O^\a)$
is BRST invariant by itself, one can check from \brstx\ that
$\d_Q\d_Q w_\a = -\Pi^m (\g_m\l)_\a.$ Note that such a transformation for
$w_\a$ is not inconsistent with $Q^2=0$ since $\d w_\a =
-\Pi^m (\g_m\l)_\a$ is a gauge transformation of the type discussed
in \defwgauge. So 
$$\d_Q\d_Q(w_\a \O^\a) =  (\d_Q\d_Q w_\a) \O^\a =-\Pi_m (\l\g^m\O),$$
which implies that $S_{BRST}$ of \clbrst\ is BRST-invariant.

It can be easily checked that this construction of $S_{BRST}$ 
agrees with the superparticle and superstring actions constructed
using pure spinors. For example, for the heterotic superstring
in an on-shell super-Yang-Mills background, 
\eqn\superpf{S_{GS}= S_{het} +
\int d^2 z [\p\t^\a A_\a^I + \Pi^m B_m^I] \bar J^I}
where $S_{het}$ is defined in \het, $\bar J^I$ are the right-moving
$E_8\times E_8$ or $SO(32)$ currents, $I$ is a Lie algebra index,
and $A_\a$ and $B_m$ satisfy \eom\ and \fs.
One can use \kappatr\ together with $\d \bar J^I=-ig
[A_\a,\bar J]^I$ to compute that 
$\O^\a = \bar\p\t^\a + W^{\a I}\bar J^I$ where $W^\a$ is defined
in \fs. So
\eqn\superppure{S_{BRST}
= S_{GS} +\int d^2 z ~\d_Q(w_\a \bar\p\t^\a +  w_\a W^{\a I}\bar J^I)}
$$= S_{GS} +\int d^2 z [d_\a\bar\p\t^\a + w_\a \bar\p\l^\a
 +  (d_\a  W^{\a I} + w_\a\l^\b (\nabla_\b W^\a)^I)\bar J^I]$$
$$=
\int d^2 z[\half\p x^m \bar\p x_m + p_\a\bar\p\t^\a + w_\a \bar\p\l^\a
+  (\p\t^\a A^I_\a +\Pi^m B^I_m +d_\a  W^{\a I} +\half N_{mn} F^{mn I}) 
\bar J^I],$$
which is the pure spinor version of the heterotic superstring action
in a super-Yang-Mills background.

\subsec{Open superstring and supersymmetric Born-Infeld equations}
 
Over fifteen years ago,
it was shown that one-loop conformal invariance of the bosonic open
string in an electromagnetic background implies that the background
satisfies the Born-Infeld equations, and higher-loop
conformal invariance implies higher-derivative corrections to
these equations
\ref \tseytlin{E.S. Fradkin and A.A. Tseytlin,
{\it Non-Linear Electrodynamics from Quantized Strings},
Phys. Lett. B163 (1985) 123\semi
A. Abouelsaood, C.G. Callan, C.R. Nappi and S.A. Yost,
{\it 
Open Strings in Background Gauge Fields}, Nucl. Phys. B280 (1987) 599\semi
O.D. Andreev and A.A. Tseytlin, {\it Two-Loop Beta Function in the Open
String Sigma Model and Equivalence with String Effective
Equations of Motion}, Mod. Phys. Lett. A3 (1988) 1349.}.
However, because of problems with describing
fermionic backgrounds, this result was generalized only
to the bosonic sector of supersymmetric Born-Infeld theory
using the Ramond-Neveu-Schwarz formalism of the open superstring \ref\town
{E. Bergshoeff, E. Sezgin, C.N. Pope and P.K. Townsend,
{\it The Born-Infeld Action from Conformal Invariance of the Open
Superstring}, Phys. Lett. B188 (1987) 70\semi
O.D. Andreev and A.A. Tseytlin, {\it Partition Function Representation
for the Open Superstring Effective Action: Cancellation of Mobius
Infinities and Derivative Corrections to Born-Infeld Lagrangian},
Nucl. Phys. B311 (1988) 205.}.
Although fermionic backgrounds can be classically described using the
Green-Schwarz formalism of the superstring, quantization
problems have prevented computation of the equations implied
by one-loop or higher-loop conformal invariance. Nevertheless,
it has been argued that $\kappa$-symmetry of the classical
Green-Schwarz superstring action in an abelian background implies
the abelian supersymmetric Born-Infeld equations for the background
\ref\howes{C.S. Chu, P.S. Howe and E. Sezgin, {\it Strings and
D-branes with Boundaries}, Phys. Lett. B428 (1998) 59, hep-th/9801202.}
\ref\sbi{S.F. Kerstan, {\it
Supersymmetric Born-Infeld from the D9-Brane}, Class. Quant.
Grav. 19 (2002) 4525, hep-th/0204225.}.

Using the pure spinor description of the 
superstring, physical states are defined using
the left and right-moving BRST charges
\eqn\brstone{Q=\int d\s (\l^\a d_\a) \quad{\rm and} \quad
\hat Q=\int d\s (\hat\l^\a \hat d_\a)}
where $d_\a$ and $\hat d_\a$
are left and right-moving worldsheet variables for the N=2
D=10 supersymmetric derivatives and
$\l^\a $ and $\hat\l^\a$ are left and right-moving
pure spinor variables satisfying
\eqn\pure{\l\g^m_{\a\b}\l^\b= \hat \l^\a \g^m_{\a\b}\hat \l^\b=0}
for $m=0$ to 9.
As was shown with Vladimir Pershin,
classical BRST invariance of the
open superstring in a background implies that the
background fields satisfy the full non-linear supersymmetric Born-Infeld
equations of motion.
This was verified by
computing the boundary conditions of the open superstring worldsheet
variables in the presence of the
background and showing that
the left and right-moving BRST currents satisfy
\eqn\equali{\l^\a d_\a = \hat\l^\a
\hat d_\a}
on the boundary if and only if the background fields satisfy
the supersymmetric Born-Infeld equations of motion.
Since $\l^\a d_\a$ is left-moving and
$\hat\l^\a \hat d_\a$ is right-moving,
${\p\over{\p\tau}} (Q+\hat Q)=
\int d\s ~{\p\over{\p\sigma}}(\l^\a d_\a -\hat\l^\a \hat d_\a).$
So \equali\ implies that
classical BRST
invariance is preserved in the presence of the open superstring
background. Although similar results
can be obtained using
$\kappa$-symmetry in the classical Green-Schwarz formalism, this
pure spinor approach has the advantage of allowing the computation of
higher-derivative
corrections through the requirement of quantum BRST invariance.

The first step in computing the equations implied
by classical BRST invariance is to determine
the appropriate boundary conditions for the open superstring worldsheet
variables in the presence of the background. Recall that for
the bosonic string in an electromagnetic background, the
Neumann boundary conditions ${\p\over{\p\sigma}} x^m =0$
are modified to
\eqn\bosboundary{{\p\over{\p\sigma}} x^m = F^{mn} \dot x^n}
where $F^{mn}$ is
the electromagnetic field strength. For the bosonic string,
these modified boundary conditions do not affect classical
BRST invariance
since \bosboundary\ together with $F^{mn}=-F^{nm}$
implies that
the left-moving stress-tensor $T= \half\p x^m \p x_m$
remains equal to
the right-moving stress-tensor $\hat T= \half\bar\p x^m\bar\p x_m$
on the boundary
where $\p = {\p\over{\p\tau}}+
{\p\over{\p\sigma}}$ and
$\bar\p = {\p\over{\p\tau}}-
{\p\over{\p\sigma}}$. So by defining the left and
right-moving reparameterization ghosts to satisfy
$c=\hat c$ and $b=\hat b$ on the boundary, one is guaranteed
that the left and right-moving BRST currents coincide on the boundary
in the presence of the background.

However, for the superstring using the pure spinor formalism,
the boundary conditions on the worldsheet variables
in the presence of a background do not
automatically imply that the left and right-moving BRST
currents coincide on the boundary. As
will be reviewed here, $\l^\a d_\a = \hat\l^\a
\hat d_\a$ on the boundary
if and only if the background superfields satisfy
the supersymmetric Born-Infeld equations of motion.

In a background, the open superstring action using the pure
spinor description is $S=S_0 +V$ where
\eqn\freeact{
S_0=-{1\over{\a'}}
 \int\! d\tau d\s \biggl\{ \half \p x^m \bar \p x_m + p_\a \bar\p \t^\a
+  \wh p_\a \p \hth^\a + w_\a \bar\p \l^\a + \wh w_\a \p \hl^\a
\biggr\}  }
is the action in a flat background
and $V$ is the super-Maxwell
integrated vertex operator defined in \intv.
Before computing the boundary conditions on the worldsheet variables
in the presence of $V$, it is convenient to add a surface term
$S_b$ to the action such that $S= S_0 + S_b$ 
is manifestly invariant under the N=1 D=10 supersymmetry transformations
\eqn\susyone{\d\t_+^\a=\e^\a,\quad \d x^m =\half \t_+\g^m\e,\quad
\d\t_-^\a=0,}
where $\t_\pm^\a = {1\over{\sqrt{2}}}(\t^\a \pm\wh\t^\a)$. 
Note that although $S_0$ is invariant under \susyone\ using the
flat boundary conditions $\t_-^\a = \p_\sigma x^m=0$, 
it is not invariant under \susyone\ for more general boundary
conditions. However, it was shown in \born\ 
that by choosing $S_b$ appropriately,
one can make $S=S_0+S_b$ invariant under \susyone\
for arbitrary boundary conditions. Furthermore, it is convenient
to modify the vertex operator $V$ to
\eqn\vertexmod{V=
\dot \t_+^\a A_\a(x,\t_+) +\Pi_+^m  B_m(x,\t_+) + d_\a^+ W^\a(x,\t_+) 
+\half N_+^{mn} F_{mn}(x,\t^+)}
where the $+/-$ index denotes the sum/difference of left and right-moving
worldsheet variables. With this modification of $V$, the background
superfields transform covariantly under the N=1 D=10 supersymmetry
transformations of \susyone.

As was shown in \born, cancellation of the surface term equations of motion
implies that the flat boundary conditions
\eqn\flatbound{\t_-^\a = 
\Pi_-^m = d_\a^- = \l_-^\a = w_\a^-=0}
are modified in the presence of $V$ to
\eqn\bctheta{ \t_-^\a = - W^\a (x,\t_+), }
\eqn\bcrest{ \Pi_-^m
= \dot\t_+^\a (\p^m A_\a - D_\a B^m + 
\g_{\a\b}^m W^\b + {1\over 6} \g^n_{\a\b}\g_{n\g\d} W^\b W^\g \p^m
W^\d ) }
$$ {}+ \Pi_+^n (\p^m B_n - \p_n B^m) +  d_\a^+ \p^m W^\a +
{1\over 2} N_+^{nk} \p^m F_{nk} ,
$$
$$ \sqrt{2} d_\a^- = \dot\t_+^\b (D_\a A_\b + D_\b A_\a -
\g^m_{\a\b}B_m + {1\over 6} \g^n_{\a\g}\g_{n\d\l} W^\g W^\d D_\b
W^\l
$$
$$ {} + {1\over 6} \g^n_{\b\g}\g_{n\d\l} W^\g W^\d D_\a W^\l )
$$
$$ {}+ \Pi_+^m (\p_m A_\a - D_\a B_m +  \g_{m\a\b} W^\b +
{1\over 6} \g^n_{\a\b}\g_{n\g\d} W^\b W^\g \p_m W^\d )
$$
$$ {}+  d_\g^+ D_\a W^\g - {1\over 2} N_+^{mn} D_\a
F_{mn} ,
$$
$$ \l_-^\a = - {1\over 4} \l_+^b (\g_{mn})_\b{}^\a F^{mn} , \qquad
w^-_\a = {1\over 4} F^{mn} (\g_{mn})_\a{}^\b w^+_\b.$$

Using the boundary conditions of \bctheta\ and \bcrest, the
difference between the left and right-moving BRST currents on
the boundary is
$$ 2 (\l^\a d_\a - \hl^\a \wh d_\a) = $$
$$  \l_+^\a \dot\t_+^\b \Bigl[D_\a A_\b + D_\b A_\a - \g^m_{\a\b}B_m
+ {1\over 6} \g^m_{\a\g}\g_{m\d\l} W^\g W^\d D_\b W^\l + {1\over 6}
\g^m_{\b\g}\g_{m\d\l} W^\g W^\d D_\a W^\l
$$
$$ {}+ {1\over 8} (\g F)_\a{}^\k \g^m_{\k\l} W^\l (\p_m A_\b - D_\b
B_m + \g_{m\b\g} W^\g + {1\over 6} \g^n_{\b\s}\g_{n\g\d} W^\s W^\g \p_m
W^\d ) \Bigr]
$$
$$ {}+ \l_+^\a \Pi_+^m \Bigl[ \p_m A_\a - D_\a B_m + \g_{m\a\b} W^\b +
{1\over 6} \g^n_{\a\b}\g_{n\g\d} W^\b W^\g \p_m W^\d
$$
$$ {}- {1\over 8} (\g F)_\a{}^\b \g^n_{\b\g} W^\g (\p_n B_m - \p_m
B_n) \Bigr]
$$
$$ {}+ \l_+^\a d_\g^+ \Bigl[ D_\a W^\g - {1\over 4} (\g F)_\a{}^\g +
{1\over 8} (\g F)_\a{}^\b \g^n_{\b\l} W^\l \p_n W^\g \Bigr]
$$ \eqn\differ{ {} - \half \l_+^\a N_+^{mn} \Bigl[ D_\a F_{mn} +
{1\over 8} (\g F)_\a{}^\b \g^k_{\b\l} W^\l \p_k F_{mn} \Bigr] , }
where $(\g F)_\b{}^\a = F_{mn}(\g^{mn})_\b{}^\a.$

Requiring this to be zero implies the equations:
$$ D_\a A_\b + D_\b A_\a - \g^m_{\a\b}B_m + {1\over 6}
\g^m_{\a\g}\g_{\d\l} W^\g W^\d D_\b W^\l + {1\over 6} \g^m_{\b\g}\g_{m\d\l}
W^\g W^\d D_\a W^\l
$$ \eqn\BIfirst{ {} + {1\over 64} (\g F)_\a{}^\g (\g F)_\b{}^\d
\g^n_{\g\l} \g^m_{\d\s} W^\l W^\s (\p_m B_n - \p_n B_m) = 0 , }
$$ \p_m A_\a - D_\a B_m + \g_{m\a\b} W^\b + {1\over 6}
\g^n_{\a\b}\g_{n\g\d} W^\b W^\g \p_m W^\d
$$ \eqn\BIsecond{ {}- {1\over 8} (\g F)_\a{}^\b \g^n_{\b\l} W^\l (\p_n
B_m - \p_m B_n) = 0 , }
\eqn\BIthird{
D_\a W^\g - {1\over 4} (\g F)_\a{}^\g +
{1\over 8} (\g F)_\a{}^\b \g^n_{\b\l} W^\l \p_n W^\g =0,
}
\eqn\BIfourth{ \l_+^\a \l_+^\b (\g^{mn})_\b{}^\g
 \Bigl[ D_\a F_{mn} +
{1\over 8} (\g F)_\a{}^\b \g^k_{\b\l} W^\l \p_k F_{mn} \Bigr]=0 . }

As in the super-Maxwell equations of \qvsat, the contraction of
\BIfirst\ with $\g_{mnpqr}^{\a\b}$ implies the equations of motion for
$A_\a$, the contraction of \BIfirst\ with $\g_m^{\a\b}$ defines
$B_m$, the contraction of \BIsecond\ with $\g^{m\a\g}$ defines
$W^\g$, the contraction of \BIthird\
with $(\g_{rs})_\b{}^\a$ defines $F_{rs}$, and the remaining
contractions of \BIsecond\ and \BIthird\ are implied by
these equations through Bianchi identities. Note that
because of the non-linear terms in \BIfirst -\BIthird,
$W^\g$ and $F_{mn}$ are now complicated functions of the
spinor and vector field strengths constructed from the gauge fields
$A_\a$ and $B_m$.

Finally, equation
\BIfourth\ vanishes as a consequence of \BIthird\ and the
pure spinor property
\eqn\purex{ \l_+\g^m\l_+ +
{1\over 16} (\g F)_\g{}^\a (\g F)_\d{}^\b \g^m_{\a\b} \l_+^\g \l_+^\d
= \l_+\g^m\l_+ + \l_-\g^m\l_-
= \l\g^m\l + \hl\g^m\hl =0.}
To show that \BIfourth\ vanishes, it is useful to write
\BIthird\ and \BIfourth\ as
$\hat D_\a W^\g = {1\over 4} (\g F)_\a{}^\g $ and
$\l_+^\a \l_+^\b \hat D_\a \hat D_\b W^\g=0$
where
\eqn\dwidehat{\hat D_\a = D_\a + \half D_\a W^\g
\Bigl(\d^\g_\b - \half \g^n_{\b\l} W^\l \p_n W^\g \Bigr)^{-1}
(\g^r W)_\b \p_r.}
One can check that
\eqn\checkcov{\{\wh D_\a , \wh D_\b\} = (\g^m_{\a\b} +
{1\over {16}} (\g F)_\a{}^\g (\g F)_\b{}^\d \g^m_{\g\d} ) \hat \p_m}
where
\eqn\pwidehat{
\hat \p_m = \p_m + \half\p_m W^\g
\Bigl(\d^\g_\b - \half \g^n_{\b\l} W^\l \p_n W^\g \Bigr)^{-1}
(\g^r W)_\b \p_r,}
so \purex\ implies that
$\l_+^\a \l_+^\b \hat D_\a \hat D_\b W^\g=0$.

To prove that equations \BIfirst - \BIthird\ are the
abelian supersymmetric Born-Infeld equations, it was 
shown in \born\ that they are invariant under N=2 D=10 supersymmetry
where the second supersymmetry acts non-linearly on the superfields.
Except for factors of $i$ coming from different conventions for the
supersymmetry algebra, equations \BIfirst - \BIthird are
easily shown to coincide with the superspace Born-Infeld equations
(33)-(35) of reference \sbi\ which were independently derived using
the superembedding method \howes.

\subsec{Closed superstring and Type II supergravity equations}

In a curved background, the classical GS superstring action can
be written as
\eqn\iigs{S=
{1\over{4\pi \a'}}\int d^2 z (G_{MN}(Z)+ B_{MN}(Z))\p Z^M\bar \p Z^N}
where $M=[m,\mu,\wh\mu]$ are curved N=2 D=10 superspace
indices, $Z^M = [x^m,\t^\mu,\wh\t^{\wh\mu}]$, 
$\mu$ and $\wh\mu$ denote SO(9,1) spinors of opposite
chirality for the Type IIA superstring and of the same chirality
for the Type IIB superstring,
and $G_{MN}$ and $B_{MN}$ describe the background superfields.
When the background fields satisfy the Type II supergravity equations
of motion, the action of \iigs\ is invariant under 
$\k$-symmetry. However, because of quantization problems,
it is not known how to use the action of \iigs\
to compute $\a'$ corrections to the supergravity equations.
This is an important question since it is not yet understood
how the superspace structure of Type II supergravity equations
is modified by these $\a'$ corrections.

As will be reviewed in this subsection, an analogous action can
be constructed using the pure spinor description of the Type II
superstring in a curved background. As was shown with Paul Howe
in \withhowe, classical BRST invariance of this action implies the
Type II supergravity equations and quantum BRST invariance is
expected to imply $\a'$ corrections to these equations.
Except for the Fradkin-Tseytlin term which couples the dilaton
to worldsheet curvature, the Type II sigma model
action in a curved background 
can be constructed by
adding the massless integrated closed superstring vertex operator 
of \closedvertex\
to the flat action of \freeact, and then covariantizing
with respect to N=2 D=10 super-reparameterization invariance.
Alternatively, one can consider the most general action constructed
from the closed superstring worldsheet variables which
is classically invariant under
worldsheet conformal transformations.

Using
the worldsheet variables of the previous subsection, the
Type II sigma model action is defined as
\eqn\IIsm{
S= {1\over {2\pi\a'}} \int d^2 z [\half 
(G_{MN}(Z)+B_{MN}(Z)) \p Z^M \bar\p Z^N
+ P^{\a\bh}(Z) d_\a \hd_\bh}
$$
+ E_M^\a(Z) d_\a \bar\p Z^M +
E_M^\ah(Z) \hd_\ah \p Z^M
+ \Omega_{M\a}{}^\b(Z) \l^\a w_\b \bar\p Z^M
+ \hat\Omega_{M\ah}{}^\bh(Z)\hl^\ah \hat w_\bh \p Z^M
$$
$$+ C_{\a}^{\b\gh}(Z) \l^\a w_\b \hd_\gh
+\hat C_{\ah}^{\bh\g}(Z) \hl^\ah \hat w_\bh  d_\g +
S_{\a\gh}^{\b\dh}(Z) \l^\a w_\b \hl^\gh \hat w_\dh ~+\half\a' \Phi(Z) r ]
+ S_\l + S_{\hl}$$
where $M=(m,\mu,\hat\mu)$ are curved superspace indices, $Z^M=(x^m,\t^\mu,
\hht^{\hat\mu})$,
$A=(a,\a,\ah)$
are tangent superspace indices, $S_\l$ and $S_{\hl}$ are the flat actions
for the pure spinor variables,
$r$ is the worldsheet
curvature, and
$[G_{MN}=\eta_{cd} E_M^c E_N^d,B_{MN},$
$ E_M^\a, E_M^\ah,\O_{M\a}{}^\b,\hat \Omega_{M\ah}{}^\bh,$
$P^{\a\bh}, C_{\a}^{\b\gh},\hat C_{\ah}^{\bh\g}, S_{\a\gh}^{\b\dh},\Phi]$
are the background superfields.
Note that $d_\a$ and $\hd_\ah$
can be treated as independent variables in \IIsm\
since $p_\a$ and $\hat p_\ah$ do not appear explicitly.

If the Fradkin-Tseytlin term $\int d^2 z \Phi(Z)r$ is omitted,
\IIsm\ is the most general action with classical worldsheet
conformal invariance and zero (left,right)-moving ghost number
which can be constructed
from the Type II worldsheet variables. Note that $d_\a$ carries
conformal weight $(1,0)$, $\hat d_\ah$ carries conformal weight
$(0,1)$, $\l^\a$ carries ghost number $(1,0)$ and
conformal weight $(0,0)$,
$\hl^\ah$ carries ghost number $(0,1)$ and
conformal weight $(0,0)$,
$w_\a$ carries ghost number $(-1,0)$ and
conformal weight $(1,0)$, and
$\hat w_\ah$ carries ghost number $(0,-1)$ and
conformal weight $(0,1)$.
Since
$w_\a$ and $\hat w_\ah$ can only appear in combinations which commute with
the pure spinor constraints, the background superfields
must satisfy
\eqn\conbaII{(\g^{bcde})_\b^\a
\Omega_{M\a}{}^\b =
(\g^{bcde})_\b^\a
\hat\Omega_{M\ah}{}^\bh =0,}
$$
(\g^{bcde})_\b^\a C_{\a}^{\b\gh}
=(\g^{bcde})_\bh^\ah \hat C_{\ah}^{\bh\g} =
(\g^{bcde})_\b^\a S_{\a\gh}^{\b\dh} =
(\g^{bcde})_\dh^\gh S_{\a\gh}^{\b\dh} = 0,$$
and the different components of the spin connections will be defined as
\eqn\OmegaII{\Omega_{M\a}{}^\b= \Omega_M^{(s)} \d_\a^\b +\half
\Omega_M^{cd} (\g_{cd})_\a{}^\b,
\quad
\hat\Omega_{M\ah}{}^\bh=\hat\Omega_M^{(s)} \d_\ah^\bh +\half
\hat\Omega_M^{cd} (\g_{cd})_\ah{}^\bh.}

Although the background superfields appearing in \IIsm\ look
unconventional, they all have physical interpretations. The superfields
$E_M{}^A$, $B_{MN}$ and $\Phi$ are the supervielbein, two-form potential
and dilaton superfields, $P^{\a\bh}$ is the superfield whose lowest
components are the Type II Ramond-Ramond field strengths, and
the superfields
$C_\a^{\b\gh}=C^\gh \d_\a^\b +\half C^{\gh ab}(\g_{ab})_\a^\b$ and
$\hat C_\ah^{\bh\g}=\hat C^\g \d_\ah^\bh +\half
 \hat C^{\g ab}(\g_{ab})_\ah^\bh$
are related to the N=2 D=10 dilatino and gravitino field strengths.
Unlike the GS sigma model of \iigs\ where the spinor supervierbein
is absent, the action of \IIsm\ contains $E_M^\a$ and
$E_m^\ah$. This means that the action is invariant under 
two sets of
local Lorentz and scale transformations which act independently
on the unhatted and hatted spinor indices. 
One therefore has two
independent sets of spin connections and scale connections,
$(\Omega_M^{(s)},\Omega_M^{ab})$
and $(\hat\Omega_M^{(s)},\hat\Omega_M^{ab})$, which appear explicitly
in the Type II sigma model action. 
Under the two types of local Lorentz and scale transformations,
\eqn\localtwo{
\d E^\a_M =\Sigma^\a_\b E^\b_M,\quad
\d E^\ah_M =\hat\Sigma^\ah_\bh E^\bh_M,\quad \d d_\a =
-\Sigma^\b_\a d_\b, \quad \d \hat d_\ah=-\hat\Sigma_\ah^\bh \hat d_\bh, }
$$
\d \Omega_{M\a}{}^\b = \p_M\Sigma_\a^\b +
\Sigma^\g_\a \Omega_{M\g}{}^\b -
\Sigma^\b_\g \Omega_{M\a}{}^\g,
\quad \d \hat\Omega_{M\ah}{}^\bh = \p_M\hat\Sigma_\ah^\bh +
\hat\Sigma^\gh_\ah \hat\Omega_{M\gh}{}^\bh -
\hat\Sigma^\bh_\gh \hat\Omega_{M\ah}{}^\gh,$$
$$\d \l^\a =
\Sigma^\a_\g \l^\g,\quad
\d w_\a =
-\Sigma^\g_\a w_\g, \quad
\d \hl^\ah =
\hat\Sigma^\ah_\gh \hl^\gh,\quad
\d \hat w_\ah =
-\hat\Sigma^\gh_\ah w_\gh,$$
where $\Sigma_\a^\b
= \Sigma^{(s)}\d_\a^\b +\half \Sigma^{bc}(\g_{bc})_\a{}^\b$,
$\hat\Sigma_\ah^\bh
= \hat\Sigma^{(s)}\d_\ah^\bh +\half \hat\Sigma^{bc}(\g_{bc})_\ah{}^\bh$,
$\Sigma^{bc}$ and $\hat\Sigma^{bc}$
parameterize independent local Lorentz transformations
on the unhatted and hatted spinor indices, $\Sigma^{(s)}$
and $\hat\Sigma^{(s)}$
parameterize independent local scale transformations on the unhatted and
hatted spinor indices, and the background superfields $[P^{\a\ah},
C_\a^{\b\gh},\hat C_{\ah}^{\bh\g}, S_{\a\gh}^{\b\dh}]$ transform
according to their spinor indices.

Finally, the background superfields
$S_{\a\gh}^{\b\dh}$ appearing in \IIsm\ are related to
curvatures constructed from the spin and scale connections.
Note that a similar relation occurs in the Type II RNS sigma model action
which contains the terms
\eqn\RNSsm{
{1\over{4\pi\a'}}\int d^2 z(\Omega_m^{ab}(x)\psi_a\psi_b \bar\p x^m
+\hat\Omega_m^{ab}(x)\bar\psi_a\bar\psi_b \p x^m +
S_{abcd}(x) \psi^a\psi^b\bar\psi^c\bar\psi^d)}
where $\psi^a= e^a_m(x)\psi^m$, $\bar\psi^a= e^a_m(x)\bar\psi^m$,
and $e_m^a(x)$ is the target-space vielbein.

It is important to note that the Fradkin-Tseytlin term
$\int d^2 z \Phi(Z)r$ is absent from the GS action of \iigs\ 
since it breaks $\k$-symmetry. However, as was argued in \withhowe,
this term is necessary in the pure spinor description in order
to preserve quantum BRST invariance and conformal invariance. The
presence of this term can also be justified by the coupling constant
dependence $e^{(2g-2)\phi}$ of genus $g$ scattering amplitudes.

As was shown in \withhowe, classical BRST invariance
of \IIsm\ implies that the background superfields satisfy
the Type II supergravity equations. For the action of \IIsm\
to be BRST invariant, it is necessary that the BRST currents
are nilpotent and holomorphic, i.e. that
$\{Q,Q\}=
\{\hat Q,\hat Q\}=
\{ Q,\hat Q\}=0$ and that
$\bar\p(\l^\a d_\a) = \p(\hat\l^\ah \hat d_\ah)=0$. 

To analyze the conditions implied by nilpotency,
it is convenient to use
the canonical momenta
$P_M = \p L/ \p (\p_0 Z^M)$ to write
\eqn\defnilII{d_\a = E_\a^M [P_M + \half
B_{M N} (\p Z^N -\bar\p Z^N) - \O_{M\b}{}^\g \l^\b w_\g
- \hO_{M\bh}{}^\gh \hl^\bh \hat w_\gh],}
$$\hd_\ah = E_\ah^M [P_M + \half
B_{M N} (\p Z^N -\bar\p Z^N) - \O_{M\b}{}^\g \l^\b w_\g
- \hO_{M\bh}{}^\gh \hl^\bh \hat w_\gh].$$

Using the canonical commutation relations
$$[P_M,Z^N\}=-i\d_M^N,\quad [w_\a, \l^\b] = -i\d_\a^\b,\quad
[\hat w_\ah, \hat \l^\bh] = -i\d_\ah^\bh,$$
one finds that
$$\{Q,Q\}= \oint \l^\a\l^\b [T_{\a\b}{}^C D_C +
\half (\p Z^N-\bar\p Z^N)H_{\a\b N}
- R_{\a\b\g}{}^\d \l^\g w_\d - \hat R_{\a\b\gh}{}^\dh \hl^\gh \hat w_\dh],$$
$$\{\hat Q,\hat Q\}= \oint \hl^\ah\hl^\bh [T_{\ah\bh}{}^C D_C +
\half (\p Z^N-\bar\p Z^N)H_{\ah\bh N}
- R_{\ah\bh\g}{}^\d
\l^\g w_\d - \hat R_{\ah\bh\gh}{}^\dh \hl^\gh \hat w_\dh],$$
$$\{Q,\hat Q\}= \oint \l^\a\hl^\bh [T_{\a\bh}{}^C D_C +
\half (\p Z^N-\bar\p Z^N)H_{\a\bh N}
- R_{\a\bh\g}{}^\d \l^\g w_\d - \hat R_{\a\bh\gh}{}^\dh \hl^\gh \hat w_\dh],$$
where $D_C = E_C^M (P_M - \O_{M\a}{}^\b \l^\a w_\b
 - \hat \O_{M\ah}{}^\bh \hl^\ah \hat w_\bh )$,
$T_{AB}{}^\a$ and $R_{AB\b}{}^\g$ are defined
using the $\Omega_{M\b}{}^\g$ spin connection,
and $T_{AB}{}^\ah$ and $\hat R_{AB\bh}{}^\gh$ are defined
using the $\hat\Omega_{M\bh}{}^\gh$ spin connection.

So nilpotency of $Q$ and $\hat Q$ implies that
\eqn\nilII{\l^\a \l^\b T_{\a\b}{}^C = \l^\a \l^\b H_{\a\b B}= \l^\a\l^\b
\hat R_{\a\b\gh}{}^\dh = \l^\a\l^\b\l^\g R_{\a\b\g}{}^\d =0,}
$$\hl^\ah \hl^\bh T_{\ah\bh}{}^C 
= \hl^\ah \hl^\bh H_{\ah\bh B}= \hl^\ah\hl^\bh
\hat R_{\ah\bh\g}{}^\d = \hl^\a\hl^\b\hl^\gh R_{\ah\bh\gh}{}^\dh =0,$$
$$
\l^\a \hl^\bh
T_{\a\bh}{}^C = 
\l^\a \hl^\bh
H_{\a\bh B}= \l^\a \l^\b R_{\a\gh\b}{}^\d=
\hat\l^\ah \hat\l^\bh \hat R_{\g\ah\bh}{}^\dh= 0,$$
for any pure spinors $\l^\a$ and $\hl^\ah$.
One can easily check that the nilpotency constraints on $R_{ABC}{}^D$
in \nilII\ are implied through Bianchi identities by the nilpotency constraints
on $T_{AB}{}^C$. Since $\l^\a$ and $\hl^\ah$ are independent pure
spinors, the remaining constraints imply that 
\eqn\pureIIintrotwo{(\g_{mnpqr})^{\a\b} T_{\a\b}{}^C=
(\g_{mnpqr})^{\ah\bh} T_{\ah\bh}{}^C=
 T_{\a\bh}{}^C= 0,}
$$(\g_{mnpqr})^{\a\b} H_{\a\b C }=
(\g_{mnpqr})^{\ah\bh} H_{\ah\bh C }=
H_{\a\bh C }= 0$$   
for any self-dual five-form direction $mnpqr$.

As was shown in \withhowe, the constraints of \pureIIintrotwo\
can be interpreted as Type II pure spinor integrability conditions and
imply
all the essential Type II supergravity constraints. Furthermore,
it was shown in \withhowe\ that the remaining conventional
Type II supergravity constraints are implied by the holomorphicity
conditions that
$\bar\p(\l^\a d_\a)=\p(\hl^\ah \hd_\ah)=0$.

\subsec{Superstring in $AdS_5\times S^5$ background and Penrose limit}

In this subsection, a quantizable action will be constructed for
the superstring in an $AdS_5\times S^5$ background with
Ramond-Ramond flux \cov\ref\osvads{N. Berkovits and
O. Chand\'{\i}a, {\it Superstring Vertex Operators in an $AdS_5\times S^5$
Background}, Nucl. Phys. B596 (2001) 185,
hep-th/0009168.}
and its Penrose limit \ref\penrose{N. Berkovits,
{\it Conformal Field Theory for the Superstring in
a Ramond-Ramond Plane Wave Background}, JHEP 0204 (2002) 037, 
hep-th/0203248.}.
Since the action is quantizable, one can in principle compute
vertex operators and scattering amplitudes in this background which
would be very useful for testing the Maldacena conjecture.
However, because of the complicated form of the action, only
the simplest vertex operators \ref\vertads
{L. Dolan and E. Witten, {\it Vertex Operators for $AdS_3$ Background
with Ramond-Ramond Flux}, JHEP 9911 (1999) 003, hep-th/9910205.}
\osvads\
and scattering amplitudes \ref\ampads{K. Bobkov and L. Dolan,
{\it Three Graviton Amplitude in Berkovits-Vafa-Witten Variables},
Phys. Lett. B537 (2002) 155, hep-th/0201027\semi
G. Trivedi, {\it Correlation Functions in Berkovits' Pure
Spinor Formulation}, hep-th/0205217.} have
so far been computed. Nevertheless, it has been proven that the action
in an $AdS_5\times S^5$ background is conformally invariant up
to one-loop order 
\ref\bz
{N. Berkovits, M. Bershadsky, 
T. Hauer, S. Zhukov and B. Zwiebach, {\it Superstring Theory on
$AdS_2\times S^2$ as a Coset Supermanifold}, Nucl. Phys. B567 (2000) 61, 
hep-th/9907200.}
\ref\bren{B.C. Vallilo, private communication.}, 
and that the action for the Penrose limit plane
wave background is exactly conformally invariant \penrose.

The action in these backgrounds
can be obtained by either plugging in the appropriate
background fields into the Type IIB sigma model action of \IIsm\
or by requiring that the sigma model has the desired target-space
isometries and is BRST invariant. Except for the contribution
of the pure spinor ghosts, the $AdS_5\times S^5$ action is
a direct generalization of the $AdS_3\times S^3$ and
$AdS_2\times S^2$ actions which were constructed 
with the collaboration of Cumrun Vafa and Edward Witten in \ref\witten
{N. Berkovits, C. Vafa
and E. Witten, {\it Conformal Field Theory of AdS Background
with Ramond-Ramond Flux}, JHEP 9903 (1999) 018, hep-th/9902098.}, and
with the collaboration
of Michael Bershadsky, Tamas Hauer, Slava Zhukov and Barton
Zwiebach in \bz.

In either the $AdS_5\times S^5$ background with R-R flux
or its
corresponding plane wave limit,
the worldsheet action using the pure spinor description is  
\eqn\altact{{\cal S} = {\cal S}_{GS} +
\int d^2 z (d_\a \overline L^\a + \hat d_\ah L^\ah -{1\over 2}
 d_\a \hat d_\bh F^{\a\bh})
+ {\cal S}_{ghost}}
where
$F^{\a\bh}=
 {1\over{120}} F^{m_1 ... m_5}
(\g_{m_1 ... m_5})^{\a\bh}$ 
is the constant five-form self-dual
Ramond-Ramond flux. For the $AdS_5\times S^5$ background,
$F^{\a\bh}$ is an invertible $16\times 16$ matrix, whereas for
its Penrose limit, $F^{\a\bh}$ is not invertible and has rank 8.

The first term
${\cal S}_{GS}$ in \altact\ is the standard covariant GS action
\eqn\covgs{{\cal S}_{GS}=\int d^2 z [\half \eta_{mn} L^m \overline L^n +
\int dy \e^{IJK} (\g_{m\a\b}L^m_I L^\a_J L^\b_K +
\g_{m\ah\bh} L^m_I L^\ah_J L^\bh_K)]}
where $L^M$ and $\bar L^M$ are defined 
using the Metsaev-Tseytlin currents
\ref\metsaevtseytlin{R.R. Metsaev and A.A. Tseytlin, {\it Type
IIB Superstring Action in $AdS_5\times S^5$ Background},
Nucl. Phys. B533 (1998) 109, hep-th/9805028.}\ref\metsaev{R.R Metsaev,
{\it Type IIB Green-Schwarz Superstring in Plane Wave Ramond-Ramond
Background}, Nucl. Phys. B625 (2002) 70, hep-th/0112044.}
\eqn\current{G^{-1} \p G = P_m L^m + Q_\a L^\a +  Q_\ah L^\ah
+ \half J_{mn} L^{mn},}
$$G^{-1} \overline\p G
= P_m \overline L^m + Q_\a \overline L^\a +  Q_\ah \overline L^\ah
+ \half J_{mn} \overline L^{mn},$$
$G(x^m,\t^\a, \hat\t^\ah)= \exp(x^m P_m +
\t^\a Q_\a + \hat\t^\ah  Q_\ah)$ takes values in a coset
supergroup,
$[x^m, \t^\a,\hat\t^\ah]$ are
$N=2$ $D=10$ superspace variables
with
$m=0$ to 9 and $[\a,\ah]=1$ to 16,
the generators
$[P_m,Q_\a, Q_\ah, J_{mn}]$ form a super-Lie algebra with the
commutation relations
\eqn\commrel{[P^m,P^n]= \half
R^{mnpq} J_{pq}, \quad  \{Q_\a,Q_\b\}=2\g^m_{\a\b} P_m,
\quad
\{Q_\ah,Q_\bh\}=2\g^m_{\ah\bh} P_m,}
$$
[Q_\a, P^m] = \g^m_{\a\b} F^{\b\gh} Q_\gh,\quad
[Q_\ah, P^m] = -\g^m_{\ah\bh} F^{\g\bh} Q_\g,\quad
\{Q_\a, Q_\gh\}= \half J_{[mn]} \g^m_{\a\b} F^{\b\dh} \g^n_{\dh\gh},$$
$J_{mn}$ generate the usual Lorentz algebra,
$R^{mnpq}$ is the constant
spacetime curvature tensor which is related to $F^{\a\bh}$
by the identity
\eqn\curvid{R^{mnpq}(\g_{pq})_\a^\b = \g^m_{\a\g} F^{\g\dh} \g^n_{\dh\kh}
F^{\b\kh} -
\g^n_{\a\g} F^{\g\dh} \g^m_{\dh\kh}
F^{\b\kh},}
and
$\int dy \e^{IJK} (\g_{m\a\b}L^m_I L^\a_J L^\b_K +
\g_{m\ah\bh} L^m_I L^\ah_J L^\bh_K)$ is the
Wess-Zumino term which is constructed such that ${\cal S}_{GS}$
is invariant under $\kappa$-symmetry.

Under $G\to\Omega G H$ for global $\Omega$ and local $H$,
the currents $G^{-1}\p G$
are invariant up to a tangent-space Lorentz rotation using the
standard coset construction where
$[P_m,Q_\a,Q_{\hat\a}, J_{mn}]$ are the generators in $\Omega$ and
$J_{mn}$ are the generators in $H$.
Since the action is constructed from
Lorentz-invariant combinations of currents, it is therefore
invariant under the global target-space isometries generated by
$[P_m,Q_\a,Q_{\hat\a}, J_{mn}]$.
Note that because the R-R field-strength is self-dual,
only $20$ of the $45$ Lorentz generators $J_{mn}$ appear
in \commrel. So only $20$ of the $L^{mn}$ currents are nonzero in
\current. For the $AdS_5\times S^5$ background, these are the
$SO(4,1)\times SO(5)$ currents $L^{ab}$ and $L^{a'b'}$ for
$a,b=0$ to 4 and $a',b'=5$ to 9. And for the plane wave background,
these are the currents $L^{jk}$, $L^{j'k'}$, $L^{+j}$ and $L^{+j'}$
for $j.k=1$ to 4 and $j',k'=5$ to 8. 

The terms $d_\a \overline L^\a$ and $\hat d_\ah L^\ah$ in \altact\ break
kappa symmetry but allow quantization since they imply non-vanishing
propagators for $\t^\a$ and $\hat\t^\ah$. And the term
$-\half
d_\a \hat d_\bh F^{\a\bh}$ comes from the R-R vertex operator and implies
that certain components of $d_\a$ and $\hat d_\bh$ are auxiliary fields.
Finally, ${\cal S}_{ghost}$ describes the action for the worldsheet ghosts
which
is non-trivial since the pure spinors transform
under Lorentz transformations and therefore couple through their
Lorentz currents to the spacetime
connection and curvature. This ghost action
is
\eqn\ghostact{{\cal S}_{ghost}=
\int d^2 z [{\cal L}_{ghost}^{flat} + \half N_{mn} \overline L^{mn}
+ \half\hat N_{mn} L^{mn} +{1\over 4} N_{mn}\hat N_{pq}
R^{mnpq}]}
where ${\cal L}_{ghost}^{flat}$ is the free Lagrangian in a flat background
for the left and right-moving worldsheet ghosts $(\l^\a,w_\a)$
and $(\hat\l^\ah,\hat w_\ah)$, 
$N_{mn}= \half \l\g_{mn}w$ and
$\hat N_{mn}= \half \hat\l\g_{mn}
\hat w$ are their left and right-moving
Lorentz currents, and $R^{mnpq}$ is the target-space curvature tensor defined
in \curvid.
Note that ${\cal S}_{ghost}$ is invariant under local tangent-space
Lorentz rotations, which is necessary for the action to be well-defined
on the coset superspace described by $G(x,\t,\hat\t)$.

To check that the action is classically BRST invariant,
i.e. that $\bar\p(\l^\a d_\a)=\p(\hat\l^\ah \hat d_\ah)=0$,
it is useful to first
compute the equations of motion for $d_\a$ and $\hat d_\ah$.
Suppose one varies $Z^M=[x^m,\t^\a,\hat\t^\ah]$ such that
$E^\a_M \d Z^M=\rho^\a$,
$E^\ah_M \d Z^M= \overline\rho^\ah$, and $E^m_M \d Z^M=0$ where
$L^\a= E^\a_M \p Z^M$,
$L^\ah= E^\ah_M \p Z^M$, $L^m = E^m_M \p Z^M$,
and $[L^\a,L^\ah,L^m]$ are defined in \current. Then the covariant
GS action ${\cal S}_{GS}$ transforms as
\eqn\transgs{\d {\cal S}_{GS}= 2\rho^\a L^m \g_{m\a\b}\overline L^\b +
2\overline\rho^\ah \overline L^m \g_{m\ah\bh} L^\bh.} The transformation of
\transgs\ is related to kappa symmetry since when $\rho^\a =
\kappa_\b L^m \g_m^{\a\b}$ and
$\overline\rho^\ah =
\overline\kappa_\bh L^m \g_m^{\ah\bh}$, $\d {\cal S}_{GS}$ is proportional
to the Virasoro constraints $\eta_{mn} L^m L^n$ and
$\eta_{mn}\overline L^m \overline L^n$.

Furthermore, the commutation relations of \commrel\ imply that
\eqn\transcur{\d L^\a =\p\rho^\a + {1\over 4}
(\g^{mn})^\a_\b L_{mn} \rho^\b
+F^{\a\bh}\g^m_{\bh\gh} L_m  \overline\rho^\gh,}
$$\d L^\ah =\p\overline\rho^\ah + {1\over 4}
(\g^{mn})^\ah_\bh L_{mn} \overline\rho^\bh
-F^{\b\ah}\g^m_{\b\g} L_m \rho^\g,$$
$$\d L^{mn} = (\g^{[m} F \g^{n]})_{\b\gh}\rho^\b L^\gh
+(\g^{[m} F \g^{n]})_{\b\gh} L^\b \overline\rho^\gh$$
where
$(\g^{[m} F \g^{n]})_{\a\dh}=\half( \g^m_{\a\b} F^{\b\gh} \g^n_{\gh\dh}-
\g^n_{\a\b} F^{\b\gh} \g^m_{\gh\dh}).$

So by varying $\rho^\a$ and $\overline\rho^\ah$, one obtains the equations
of motion
\eqn\motiond{\overline\p d_\a= 2\g^m_{\a\b} L_m \overline L^\b + {1\over 4}
d_\b (\g_{mn})_\a^\b
\overline L^{mn} - \hat d_\bh F^{\g\bh} \g^m_{\g\a} L_m
 +\half
(\g^{[m} F \g^{n]})_{\a\gh}(N_{mn} \overline L^\gh +\hat N_{mn} L^\gh),}
$$\p\hat d_\ah=2 \g^m_{\ah\bh} \overline L_m L^\bh + {1\over 4}
\hat d_\bh (\g_{mn})_\ah^\bh
L^{mn} +  d_\b F^{\b\gh} \g^m_{\gh\ah} \overline L_m
 -\half
(\g^{[m} F \g^{n]})_{\g\ah}(N_{mn} \overline L^\g +\hat N_{mn} L^\g).$$

Plugging into \motiond\ the equations of motion $\overline L^\a=\half F^{\a\bh}
\hat d_\bh$ and
$L^\ah= -\half F^{\b\ah}
d_\b$ which come from varying $d_\a$ and $\hat d_\ah$, one finds
\eqn\motd{\overline\nabla d_\a=
\half(\g^{[m} F \g^{n]})_{\a\gh}(N_{mn} \overline L^\gh -\half
\hat N_{mn} F^{\d\gh} d_\d),}
$$\nabla\hat d_\ah=
-\half(\g^{[m} F \g^{n]})_{\g\ah}
(\half N_{mn} F^{\g\hat\d} \hat d_{\hat\d}
+\hat N_{mn} L^\g),$$
where the spin connections in the covariantized derivatives
$\nabla$ and $\overline\nabla$ are $L^{mn}$
and $\overline L^{mn}$.

Furthermore, the equations of motion
of $\l^\a$ and $\hat \l^\ah$ coming from \ghostact\ are 
\eqn\motlambda{
\overline\nabla
\l^\a= {1\over 8} R^{mnpq} (\g_{mn})_\b^\a \l^\b \hat N_{pq},}
$$\nabla \hat\l^\ah = {1\over 8}
R^{mnpq} (\g_{pq})_\bh^\ah \hat\l^\bh N_{mn}.$$
So \motd\ and \motlambda, together with the identity
of \curvid,
imply that
\eqn\motq{\overline\p(\l^\a d_\a)= \half\l^\a
(\g^{[m} F \g^{n]})_{\a\gh} N_{mn}\overline L^\gh,}
$$\p(\hat\l^\ah \hat d_\ah)= -\half\hat\l^\ah
(\g^{[m} F \g^{n]})_{\g\ah} \hat N_{mn} L^\g.$$
Since $N_{mn}=\half (\l\g_{mn} w)$ and $\l^\a\l^\b$ is proportional
to $(\l\g^{pqrst}\l) (\g_{pqrst})^{\a\b}$, the right-hand side of \motq\
is proportional to
$\g_{mn}\g_{pqrst}\g^{[m} F \g^{n]} $. But since
$\g_m \g_{pqrst} \g^m=0$, one finds that
\eqn\propzero{\g_{mn}\g_{pqrst}\g^{[m} F \g^{n]}  =2
\g_{pqrst}\g^n F \g_n  =2\g_{pqrst} \g^n \g_{uvwxy} \g_n F^{uvwxy}=0.}
So $\overline\p(\l^\a d_\a)= \p(\hat\l^\ah \hat d_\ah)=0$ as desired.

\vskip 20pt

{\bf Acknowledgements:}
I would like to thank the ICTP members for inviting me to give
these lectures and for financial support. I would also like to thank
CNPq grant 300256/94-9, Pronex grant 66.2002/1998-9, 
and FAPESP grant 99/12763-0 for partial financial support, and all
my collaborators for their contributions.
This research was partially conducted during the period the author
was employed by the Clay Mathematics Institute as a CMI Prize Fellow.

\listrefs

\end